\pdfoutput=1
\documentclass[aps,prd,twocolumn,nofootinbib,superscriptaddress]{revtex4-2}
\usepackage{braket}
\usepackage{color}
\newcommand{\sign}{\text{sign}}  
\usepackage{graphicx}
\usepackage{amsmath}
\usepackage{amssymb}
\usepackage[utf8]{inputenc} 
\usepackage{url} 

\begin{document}

\title{Semiclassical and quantum features of the Bianchi I cosmology \\in the polymer representation}
\author{Eleonora Giovannetti}\email{eleonora.giovannetti@uniroma1.it}
\affiliation{Physics Department, “Sapienza” University of Rome, P.le Aldo Moro 5, 00185 (Roma), Italy}
\author{Giovanni Montani}\email{giovanni.montani@enea.it}
\affiliation{Physics Department, “Sapienza” University of Rome, P.le Aldo Moro 5, 00185 (Roma), Italy}\affiliation{ENEA, FSN-FUSPHY-TSM, R.C. Frascati, Via E. Fermi 45, 00044 Frascati, Italy}
\author{Silvia Schiattarella}\email{silvia.schiattarella@outlook.com}
\affiliation{Physics Department, “Sapienza” University of Rome, P.le Aldo Moro 5, 00185 (Roma), Italy}
\date{\today}	

\begin{abstract}
We analyze the Bianchi I cosmology in the presence of a massless scalar
field and describe its dynamics via a semiclassical and quantum polymer
approach by adopting three different sets of configurational variables:
the natural Ashtekar connections, a set of anisotropic volume-like coordinates and the Universe volume plus two anisotropy
coordinates (the latter two sets of variables would coincide in the case of an
isotropic Universe). In the semiclassical analysis we demonstrate that the Big Bounce clearly emerges in all the sets of variables. Moreover, when adopting the proper volume variables (i.e. when defining the Universe volume itself on the polymer lattice) we also derive the exact polymer-modified Friedmann equation for the Bianchi I model. This way, the expression of the critical energy density that includes the anisotropic energy-like contribution is obtained, demonstrating that in this set of variables the Big Bounce has a universal nature (i.e. its critical energy density has a maximum value fixed by fundamental constants and the Immirzi parameter). Finally, it is made a proposal to recover the equivalence between the dynamics in the Ashtekar set and in the anisotropic volume-like one, showing that this request is satisfied when a polymer parameter depending on the configurational coordinates is considered. Then, we apply the Arnowitt-Deser-Misner reduction of the variational principle and we quantize the system. We study the resulting Schr\"{o}dinger-like dynamics only in the Ashtekar variables and in the proper volume ones, stressing that the behavior of the Universe wave packet over time singles out common features with the semiclassical trajectories. However, we show that the standard deviation for the Universe volume operator grows with time. This is a signal of the spreading of the Bianchi I wave packet and hence of the importance of the quantum fluctuations in order to properly study the Big Bounce picture.
\end{abstract}

\maketitle
	
\section{INTRODUCTION}

One of the most relevant phenomenological implications of Loop Quantum
Gravity (LQG) \cite{Rovelli,CQG} is certainly the
emergence of a Big Bounce in the quantum dynamics of the isotropic
Universe \cite{ashtekar2003,ashtekar2005gravity,Ashtekar2006,AshtekarI,Ashtekar2008,Ashtekar2011,B,bojowald2009,Ant,M,EFG}. Despite the cosmological implementation of the formalism of the full
theory present some limitations, especially concerning the proper
assessment of the $SU(2)$ symmetry
(see \cite{C1,C2,Alesci}), the
discreteness of the geometrical operator (area and volume) spectra is at
the ground of a deep revision in the
concept of the early Universe.

Actually, the emergence of a Big Bounce and the corresponding existence of
a cut-off for the matter energy density take place in their own
evidence mostly on a semiclassical level, where the mean value of
localized wave packets follows a revised dynamics in which the singularity
is removed (for completeness, see \cite{Ashtekar2008} where for the Friedmann-Lemaitre-Robertson-Walker (FLRW) Universe it is obtained an analytically solvable model named sLQC and a full quantum analysis is performed).
On this level, the most intriguing open question concerns the specific
morphology of this semiclassical turning point in the past of the present
Universe. In fact, two possible representations are possible:
one imposing a kinematical cut-off on the minimal area element
\cite{Ashtekar2006} and the other one
assigning the same cut-off on a dynamical level, i.e. involving the cosmic
scale factor in the discrete spectrum \cite{AshtekarI,Ashtekar2008}.
In the first approach, the turning point in the past depends on the wave
packet profile as fixed at a given instant of time and also the
critical energy density can approach very large values \cite{PC,CQG}. In the dynamical reformulation, both the Big Bounce and the
critical energy  density become intrinsic features of the Universe, namely
fixed by the fundamental constants and parameters only \cite{AshtekarI}.
From the point of view of the adopted configurational variables, the
difference can be summarized in the use of the pure Ashtekar connection in
the kinematical spectrum case or the Universe volume when
an intrinsic cut-off comes out.
With respect to this dualism of possible representations, see the
discussion in \cite{EFG} where the same question is
addressed in Polymer Quantum Mechanics (PQM) by tracing a close parallelism with
Loop Quantum Cosmology (LQC). Indeed, the polymer approach \cite{ASHTEKARpol,Pol} is the most simple treatment able to provide insights on the emergence of a
bouncing cosmology without entering the
subtleties of LQG and LQC, but reliably retaining
the same information of the latter.

Here, we apply PQM to the Bianchi I model in the presence of a massless scalar field and we explore the dynamics in different sets of configurational variables by applying both the
so-called semiclassical approach and the purely quantum one
\cite{Pol}. We compare the use of the natural Ashtekar connections to the
implementation of two different sets of volume-like variables (here dubbed the  anisotropic volume-like coordinates as well as the
Universe volume plus two anisotropies),
following the LQC formulation in \cite{ashtekar2009,szulc}.
On a semiclassical level we show that, when PQM is applied to the anisotropic Universe
dynamics, the Big Bounce as an intrinsic
feature is reached only in the pure volume formulation, whereas both in the anisotropic volume-like coordinates and in the Ashtekar
connections the resulting bouncing cosmology depends on the initial conditions on the system. This result suggests that it is strictly the use of the Universe
volume coordinate that ensures an intrinsic cut-off, whereas the geometrical dimensions of the variables described as discrete on the polymer lattice is only linked to the determination of whether or not the Big Bounce exists (see \cite{EBianchiIX} where it is shown that the Bianchi I and IX models in the Misner variables are still singular in the polymer formulation). This analysis is supported by the derivation of the polymer-modified Friedmann-like equation in the pure volume formulation, from which we can analyze the expression of the total energy density at the Bounce, including the anisotropic contribution to the standard matter energy density term. Moreover, we investigate the dynamical equivalence of the three different formulations in the semiclassical polymer approach, by generalizing the analysis performed in \cite{EFG}. However, from our study comes out that only the dynamics in the Ashtekar variables and that in the anisotropic volume-like ones can be mapped, by considering the polymer parameter depending on the configurational coordinates (rather than constant) when performing the canonical change of variables.

Regarding the proper application of PQM, we analyze the evolution of the Bianchi I
cosmology only limiting our attention to the choice of the Ashtekar variables (i.e. the privileged set in LQG) and the proper volume ones (i.e. the Universe volume plus two anisotropies). We choose the matter scalar field as a clock before quantizing by means of an Arnowitt-Deser-Misner (ADM) reduction of the variational principle \cite{ADM}. Actually, the use of this Schr\"{o}dinger-like approach is justified by the need to avoid all the issues regarding the well-definition of a conserved probability density constructed using the Wheeler-DeWitt equation (in this respect, see \cite{shell,rosenstein1985probability}). In the Ashtekar variables the evolution of a localized quantum wave packet is studied by following the peaks of the probability density (i.e. the square modulus) and allows to show that there is a good coincidence between the semiclassical trajectories and the behavior of the Universe wave packet. Therefore, we can infer that the results obtained in the semiclassical sector in
terms of the Ashtekar variables remain valid in the full quantum picture. In other words, the feature of a non-universal Big Bounce depending on the initial conditions on the wave packets seems to be well-grounded on a quantum level too. Then, in the pure volume representation the quantum mean value of the volume operator is studied, showing a good consistency with the volume semiclassical trajectory. Since the Bianchi I wave packet outlines a clear spreading over time, we also compute the standard deviation of the volume operator, highlighting its non-linear behavior over time. This result points out the need of going beyond the pure semiclassical description of the Big Bounce, in view of its intrinsic quantum nature \cite{B2020}. Nonetheless, this numerical study allows us to claim the existence of the quasi-classical limit, since the relative error does not exceed the unity for a significant time interval.

The paper is structured as follows. In Sec. \ref{pol} the theory of PQM is introduced, in order to clarify the formalism at the ground of the following cosmological analysis. In Sec. \ref{iso} the main results obtained by describing the FLRW model in the polymer picture are presented, providing the analyses in the Ashtekar variables and in the volume ones into two subsections. In Sec. \ref{ham} the Hamiltonian formulation of the Bianchi I model in the Ashtekar variables is described, in order to introduce the original part of the paper developed in Secc. \ref{sem1}-\ref{quant}. In particular, in Secc. \ref{sem1} and \ref{sem2} it is solved the semiclassical polymer dynamics of the Bianchi I model in the Ashtekar variables and in the two sets of volume-like variables respectively, whereas in Sec. \ref{discussion} a proposal about the possibility to recover the equivalence between the dynamics in the three sets of variables is investigated. Furthermore, in the Ashtekar set and in the proper volume one a full quantum treatment is introduced in Sec. \ref{quant} using a Scr\"{o}dinger-like formalism. Finally, in Sec. \ref{con} some concluding remarks are commented.
		
\section{POLYMER QUANTUM MECHANICS\label{pol}}
As a first step, we introduce the polymer quantization of a system, which is a non-equivalent representation of the quantum mechanics with respect to the standard Schr\"{o}dinger one (see \cite{Pol}). This formulation is based on the assumption that one or more variables of the phase space are discretized. Hence, in this approach it is not possible to define $\hat q$ and $\hat p$ as operators at the same time. However, the power of this alternative quantum approach is the possibility to describe cut-off physics effects through a simple formalism and this is particularly useful in the cosmological setting, where the generalized coordinates are identified with the cosmic scale factors.

First of all, we introduce a set of abstract kets $\ket \mu$ with $\mu \in {\rm I\!R}$ and fix the inner product as	
\begin{equation}
\langle \mu \ket {\nu} =\delta_{\mu\nu}\,,
\end{equation}
being $\delta_{\mu\nu}$ a Kronecker delta. This procedure defines the non-separable Hilbert space $H_{poly}$ where we can define two fundamental operators: the \emph{label} operator $\hat\epsilon$, whose action on the kets is given by $\hat\epsilon\ket\mu=\mu\ket\mu$, and the \emph{shift} operator $\hat s(\lambda)$ ($\lambda\in {\rm I\!R}$), where $\hat s (\lambda)=\ket{\mu+\lambda}$. The action of $\hat s (\lambda)$ is discontinuous since the kets are orthonormal $\forall \lambda$. Therefore, no Hermitian operator could generate it by exponentiation.

Now, we suppose that the configurational coordinate $q$ has a discrete character, i.e. the position is defined on a lattice having a given spacing. The projection of the states in the $p$-polarization is ($\hbar=1$)
\begin{equation}
    \psi_{\mu}(p)=\langle p|\mu\rangle=e^{ip\mu}\,.
\end{equation}
By applying the shift operator on these states as
\begin{equation}
	\hat s(\lambda)\psi_{\mu}(p)=e^{i\lambda p}e^{i\mu p}=\psi_{(\mu+\lambda)}(p)\,,
\end{equation}
we can conclude that the operator $\hat p$ cannot be defined in a rigorous fashion, whereas the action of $\hat q$ reads as
\begin{equation}
	\hat q \psi_{\mu}(p)=-i\partial_{p}e^{i\mu p}=\mu\psi_{\mu}(p)\,,
\end{equation}
representing exactly that one of the label operator $\hat\epsilon$.

On a dynamical level we have to face the problem of defining an approximated version of the $\hat p$ operator, since the Hamiltonian $\mathcal{H}$ is a function of both $(q,p)$ as
\begin{equation}
	\mathcal{H}(q,p)={p^{2}\over 2m}+V(q)\,,
\end{equation}
where we have considered the case of a non-relativistic particle of mass $m$ in a potential $V(q)$.
In order to overcome this problem, we introduce a regular graph
\begin{equation}
	\gamma_{\mu_{0}}=\{q\in{\rm I\!R}\,|\, q=n\mu_{0}, \forall n \in \mathbb{Z}\}\,,
\end{equation}
i.e. a numerable set of equidistant points whose spacing is given by the scale $\mu_{0}$. In this way, we can restrict the action of the shift operator $e^{i\lambda p}$ by imposing $\lambda=n \mu_{0}$ in order to remain in the lattice. Then, we can use it to approximate any function of $p$ since
\begin{equation}
	\label{approxPol}
	p \approx \frac{1}{\mu_{0}} \sin \left(\mu_{0} p\right)=\frac{1}{2 i \mu_{0}}\left(e^{i \mu_{0} p}-e^{-i \mu_{0} p}\right)\,.
\end{equation}
We notice that this approximation is good for $\mu_{0}p\ll1$. Under this hypothesis one derives
\begin{equation}
	\label{phat}
	\hat{p}_{\mu_{0}}\ket{\mu_{n}}=-{i\over 2\mu_{0}}(\ket{\mu_{n+1}}-\ket{\mu_{n-1}})\,.
\end{equation}
Thus, it is possible to define a regularized operator $\hat p_{\mu_{0}}$ that depends on the scale $\mu_{0}$ and a modified version of the Hamiltonian as
\begin{equation}
	\hat{\mathcal{H}}_{\mu_{0}}:={\hat{p}^{2}_{\mu_{0}}\over 2m}+V(\hat q)\,.
\end{equation}
In what follows we will apply this same picture to configurational variables describing the Bianchi I cosmology. Each independent variable will be associated to the representation traced above for the polymer framework.
	
\section{THE ISOTROPIC UNIVERSE\label{iso}}

In this section we provide a brief analysis of the homogeneous and isotropic Universe as described by the FLRW model. The ADM line element of the FLRW Universe is given by
\begin{equation}
	\label{metricss} ds^{2}=-N(t)^{2}dt^{2}+a^{2}(t)\Big({dr^{2}\over 1-Kr^{2}}+ r^{2}d\theta^{2}+r^{2}sin^{2}\theta d\phi^{2}\Big)\,,
\end{equation}
where $N(t)$ is the lapse function, $a(t)$ the cosmic scale factor and $K=0,\pm1$ the signature of the spatial curvature. The factor $a(t)$ is the only degree of freedom available to define the dynamical properties of the Universe. Substituting the metric \eqref{metricss} in the Einstein-Hilbert action we obtain
		
\begin{equation}
	\begin{aligned}
	\label{actionaa}
	S_{\mathrm{RW}} &=\int_{t_{1}}^{t_{2}} d t\left(p_{a} \dot{a}-N \mathcal{H}_{\mathrm{RW}}\right) \\
	&=\int_{t_{1}}^{t_{2}} d t\left[p_{a} \dot{a}-N\left(-\frac{\kappa}{24 \pi^{2}} \frac{p_{a}^{2}}{a}-\frac{6 \pi^{2} K}{\kappa} a+2 \pi^{2} \rho a^{3}\right)\right],
    \end{aligned}
\end{equation}
where $\rho=\rho(a)$ is the matter energy density and $\kappa=8\pi G$ is the Einstein constant (we have set the speed of light equal to one).

To achieve a complete canonical description of the dynamics for the isotropic Universe, we consider the variations of the action above with respect to the lapse function $N$ and the conjugate variables $(a,p_{a})$. In particular, the variation of \eqref{actionaa} with respect to $N$ provides the Hamiltonian constraint for the FLRW model
\begin{equation}
	\label{HFRW}
	\mathcal{H}_{RW}=\frac{p_{a}^{2}}{a^{4}}+\frac{144 \pi^{4}}{\kappa^{2} a^{2}} K-\frac{48 \pi^{4}}{\kappa} \rho=0\,,
\end{equation}
which coincides with the Friedmann equation
\begin{equation}
	\label{eqn:Fried}
	H^{2}=\bigg({\dot a\over a}\bigg)^{2}={\kappa\rho\over 3}-{K\over a^{2}}\,.
\end{equation}
	
Finally, we notice that the dynamics of the isotropic Universe in the Hamiltonian formulation resembles that one of a one-dimensional pinpoint particle, with generalized coordinate $a$ and momentum $p_{a}$.
		
\subsection{Semiclassical polymer dynamics of the FLRW model\label{FRW}}

In order to elucidate the different features characterizing the bouncing dynamics when adopting different sets of variables in the polymer framework, we start by analyzing the semiclassical behavior of the isotropic Universe using the standard Ashtekar connection or the Universe volume. This material will provide a clear trace for the subsequent analysis on the Bianchi I cosmology.

\subsubsection{\small Analysis in the Ashtekar variables\label{FRWash}}

Firstly, we focus on the implementation of the polymer scheme to the FLRW space time for the flat Universe $(K=0)$ in the Ashtekar variables (see \cite{EFG}). In the semiclassical approach the canonical variables are restated according to the polymer formulation and the Hamiltonian dynamics is retained, providing a significant insight about the behavior of the mean values characterizing PQM and, actually, also LQC \cite{ASHTEKARpol}.

We choose the couple $(c,p)$ as the conjugate variables of the phase space, which can be expressed as functions of the scale factor as
\begin{equation} |p|=a^{2}\,, \hspace{1cm} c=\gamma \dot a\,.
\end{equation}
The scalar constraint \eqref{HFRW} in these variables reduces to
\begin{equation}
	\label{Hpc}
	\mathcal{H}=-{3\over\kappa\gamma^{2}}c^{2}\sqrt{|p|}+{p_{\phi}^{2}\over 2|p|^{3\over2}}=0\,,
\end{equation}
when a massless scalar field is included in the dynamics.
Now, in order to implement the polymer paradigm we apply the substitution (see \eqref{approxPol})
	\begin{equation} c\to{1\over\mu}\sin(\mu c)\,,
	\end{equation}
where the term $\mu$ represents the characteristic spacing of the polymer lattice on which the variable $p$ it is defined, i.e. the configurational variable that has the dimensions of an area. So, it is possible to express the Hamiltonian for the FLRW model in the polymer approach as
	\begin{equation}
	\label{HPOL} \mathcal{H}_{poly}=-{3\over \kappa\gamma^{2}}\sqrt{p}{\sin^2(\mu c)\over \mu^{2}}+{p_{\phi}^{2}\over 2|p|^{3\over2}}=0
		\end{equation}
and to write the Hamilton equations of motion for the system:
\begin{equation}
	\label{FRWeq}
\begin{cases}
	\begin{aligned}
	&\dot{p}={Nk\gamma\over3}{{\partial \mathcal{H}_{poly}}\over{\partial c}}=-{2N\over \gamma\mu}\sqrt{p}\sin(\mu c)\cos(\mu c)\\
	&\dot{c}=-{Nk\gamma\over3}{{\partial \mathcal{H}_{poly}}\over{\partial p}}={N\kappa\gamma\over{2\sqrt{p}}}\Big({\sin^{2}(\mu c)\over{\kappa\gamma^{2}\mu^{2}}}+{p_\phi^{2}\over{|p|^{3\over2}}}\Big)
	\end{aligned}
\end{cases}
\end{equation}
recalling that the commutation relations for the configurational variables are
\begin{equation} \{c, p\}={\kappa\gamma\over3}\,,
\end{equation}
where $\gamma$ is the Immirzi parameter. Here, we fix the time gauge imposing $\dot \phi:=N{{\partial H_{poly}}\over{\partial p_\phi}}=1 \Rightarrow N={|p|^{3\over2}\over{ 2 p_\phi}}\,,$ that corresponds to the choice of $\phi$ as a relational time. We notice that with the dotted variables we denote their respective $t$ derivative.	
\begin{figure}[h!]
	\centering
	\includegraphics[width=0.65\linewidth]{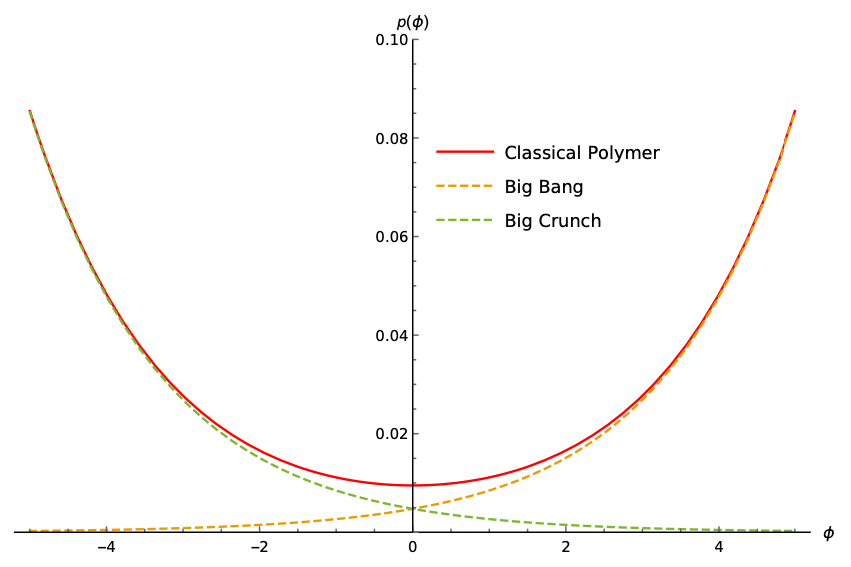}
	\caption{The polymer trajectory $p(\phi)$ (continuous line) shows that a Big Bounce regularizes the singular behavior of the classical trajectories (dotted lines) at a Planckian scale for the flat FLRW model.}
	\label{FRWbounce}
\end{figure}
	
As we can see from Fig.$\,$\ref{FRWbounce}, the polymer trajectory of the FLRW volume $V=|p|^{3\over2}$ follows the classical one (characterized by the presence of a singularity also in the Ashtekar variables) until the Universe reaches a quantum era. Here, the effects of quantum geometry due to the polymer lattice become dominant and the classical Big Bang is replaced by a quantum Big Bounce.

Using the equations \eqref{FRWeq} and the scalar constraint \eqref{HPOL} we can write the analytic expression of the Friedmann equation as
\begin{equation}
	\label{FRWFriedmann}
	H^{2}=\Big({\dot p\over 2p} \Big)^{2}={\kappa\over3}\rho\Big(1-{\rho\over\rho_{crit}} \Big)\,,
\end{equation}
where the critical energy density of the Universe (i.e. the maximum energy density which is taken at the Bounce) corresponds to

\begin{equation} \rho_{crit}=\frac{3}{\mu^2|p|}=\Big( {3\over{\kappa\gamma^{2}\mu^{2}}}\Big)^{3\over2}{1\over p_{\phi}}\,.
	\label{rhoashFLRW}
\end{equation}	
In this representation, $\rho_{crit}$ tends to $0$ when $p_{\phi}$ increases, whereas the Big Bounce approaches the Big Bang singularity when $p_{\phi}\ll1$. This result resembles that one obtained in the LQC theory in its original formulation \cite{Ashtekar2006}.
		
\subsubsection{\small Analysis in the volume variable\label{FRWV}}
	
Let us now change the variables we adopted above by introducing a new canonical set in which the generalized coordinate corresponds to the Universe volume (see \cite{EFG}), i.e.:
\begin{equation}
	v=|p|^{3\over2}=\dot a\,, \hspace{0.8cm}\eta={2c\over 3 \sqrt{|p|}}\sim{\dot a\over |a|}\,.
\end{equation}
Due to the canonicity of the transformation, the introduction of these variables conserves the algebra of the Poisson brackets. In the new set the semiclassical polymer Hamiltonian constraint rewrites as		
\begin{align}
	&\mathcal{H}_{poly}=-{27\over 4\kappa \gamma^{2}\mu^{2}}v\sin^2(\mu\eta)+{p_{\phi}^{2}\over 2v}=0\,,
\end{align}
where the polymer substitution has been implemented on the momentum conjugate to the Universe volume $v$ that is chosen as the discrete variable. In correspondence to this restated problem, the Hamilton equations take the form
\begin{equation}
		\label{eqn:FRWvolume}
	\begin{cases}
\begin{aligned}
	&\dot{v}={Nk\gamma\over3}{{\partial \mathcal{H}_{poly}}\over{\partial \eta}}=-{18N\over 4\gamma\mu}v\sin(\mu \eta)\cos(\mu \eta)\\
	&\dot{\eta}=-{Nk\gamma\over3}{{\partial \mathcal{H}_{poly}}\over{\partial v}}={N\kappa\gamma\over 3}\Big({27\sin^{2}(\mu \eta)\over{4\kappa\gamma^{2}\mu^{2}}}+{p_\phi^{2}\over 2v^{2}}\Big)
\end{aligned}
\end{cases}
\end{equation}
and can be easily solved after fixing again the time gauge $\dot{\phi}=1$, in order to provide the behavior of the Universe volume. 
\begin{figure}[h!]
	\centering
	\includegraphics[width=0.75\linewidth]{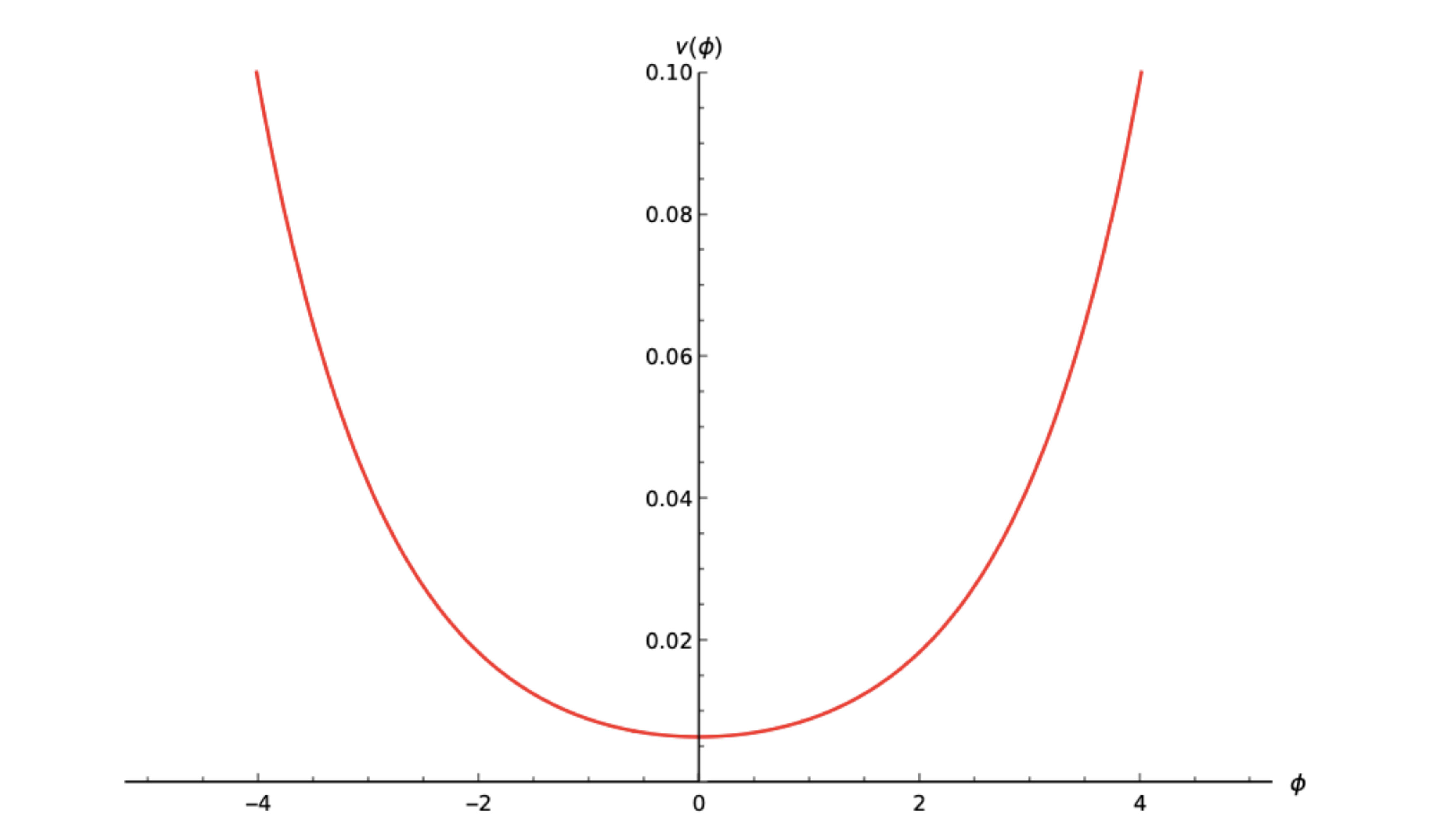}
	\caption{The polymer trajectory of the Universe volume $v$ for the FLRW model clearly shows a minimum in correspondence of the Big Bounce.}
	\label{FRWvolume}
\end{figure}

As depicted in Fig.$\,$\ref{FRWvolume} we see that the Universe has again a bouncing point in correspondence to the minimum of its volume.

To elucidate the nature of the Big Bounce, we restate the associated Friedmann equation as
\begin{equation}
	\label{FRWfrie}
	H^{2}=\Big({\dot v\over 3v} \Big)^{2}={\kappa\over3}\rho\Big(1-{\rho\over\rho_{crit}} \Big)\,,
\end{equation}
where the critical energy density is explicitly expressed by	\begin{equation} \rho_{crit}={27\over{4\kappa\gamma^{2}\mu^2}}\,.
	\label{rhovFLRW}
\end{equation}
This result shows that the energy density at which the Bounce occurs does not depend on the value assumed by the constant of motion $p_\phi$ in this representation, in contrast with the scenario obtained in the previous subsection. This analysis establishes a clear correspondence between the LQC $\bar{\mu}$-scheme \cite{AshtekarI} and the polymer approach when the volume variable is adopted.

\section{HAMILTONIAN FORMULATION OF THE BIANCHI I MODEL IN THE ASHTEKAR VARIABLES\label{ham}}
The aim of this section is to introduce some general features of the classical dynamics of the Bianchi I cosmological model as expressed in terms of the Ashtekar variables (see \cite{ashtekar2009}), before introducing the original analysis performed in the paper in the next section. The importance of studying this model resides in the legitimacy of considering more general cosmological models near the singularity with respect to the highly symmetric isotropic Universe. 

In particular, the Bianchi I model represents the simplest homogeneous but anisotropic geometry that reduces to the flat FLRW model in the isotropic limit. Its line element reads as
\begin{align}\label{BIM}{ ds^{2}=-N(t)^2 dt^2+a_1^2 dx_1^2+a_2^2 dx_2^2+a_3^2 dx_3^2}\,,
\end{align}
where $a_{1}, a_{2}, a_{3}$ are the three independent scale factors, one for each direction. The phase space of the Bianchi I model in the Ashtekar variables is six-dimensional and it is expressed through the canonical couple $(c_{i},p_j)$ defined as
\begin{equation}
	p_{i}=|\epsilon_{ijk}a_{j}a_{k}|\sign(a_{i})\,, \hspace{0.5cm}c_{i}=\gamma\dot a_{i}\,,
\end{equation}
with $i=1,2,3$ and $\{c_{i},p_j\}=\kappa\gamma\delta_{ij}$.
Starting from the metric \eqref{BIM}, we find the structure of the Bianchi I Hamiltonian constraint in the Ashtekar variables when a massless scalar field is considered, i.e.
\begin{equation}
	\label{HB}
	\mathcal{H}=-{1\over{\kappa\gamma^2V}}(c_{1}p_1c_{2}p_2+c_{1}p_1c_{3}p_3+c_{2}p_2c_{3}p_3)+{p^2_\phi\over 2 V}=0\,,
\end{equation}
with $V=\sqrt{p_1p_2p_3}$. This constraint reduces to that one of the isotropic model \eqref{Hpc} if the isotropy condition is imposed.

The classical dynamics of the system is clearly completed by the following Hamiltonian equations:
\begin{equation}
\begin{cases}
\begin{aligned}
	\label{AshBI}
	&	\dot{p_i}=Nk\gamma{{\partial \mathcal{H}}\over{\partial c_i}}=-{{p_i}\over\gamma p_{\phi}}\big(c_{j}p_{j}+c_{k}p_{k}\big)\\
	&\dot{c_i}=-Nk\gamma{{\partial \mathcal{H}}\over{\partial p_i}}={c_{i}\over{\gamma p_{\phi}}}\big(c_{j}p_{j}+c_{k}p_{k}\big)
\end{aligned}
\end{cases}
\end{equation}
for $i,j,k=1,2,3$, $i\neq j\neq k$. After imposing the time gauge $ \dot \phi=1=:N{{\partial\mathcal{H}}\over{\partial p_\phi}}\Rightarrow N={V\over p_\phi}\,,$ we can solve equations \eqref{AshBI} and the scalar constraint \eqref{HB} by assigning proper initial conditions. In particular, the initial value problem must satisfy the Hamiltonian constraint, so that the solution can be numerically provided.
\begin{figure}[h!]
	\centering
	\includegraphics[width=0.65\linewidth]{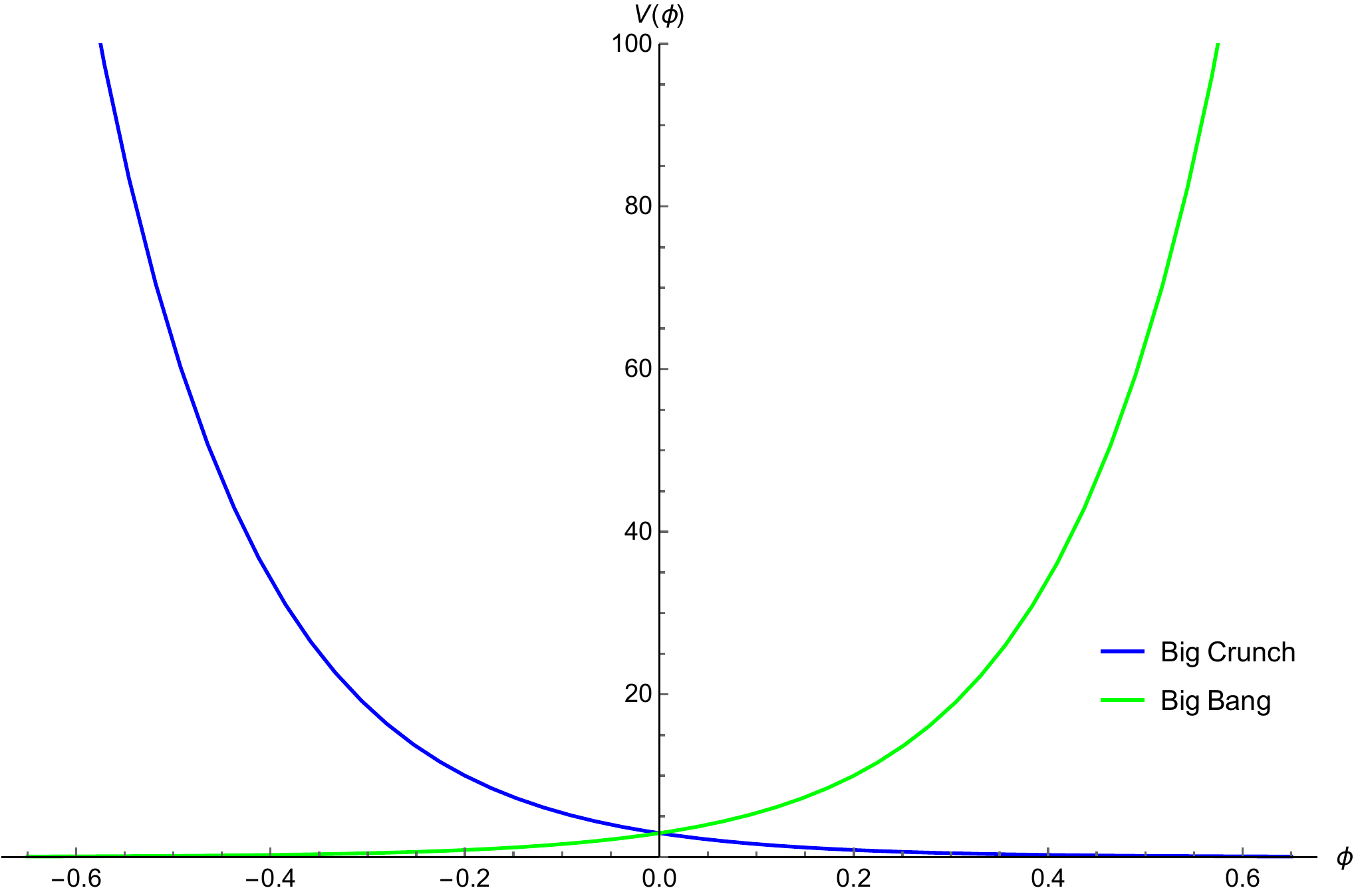}
	\caption{Trajectories of the Universe volume $V=\sqrt{p_{1}p_{2}p_{3}}$ in function of time $\phi$ in the Ashtekar variables for the Bianchi I model.}
	\label{ashtekar}
\end{figure}

Actually, once the constants of motion
\begin{equation}
	\label{eqn:integraliprimi}
	{c_{i}p_{i}}=\mathcal{K}_{i}\,,\hspace{0.5cm}p_{\phi}=\mathcal{K}_{\phi}
\end{equation}
have been identified ($i=1,2,3$), the six-equations system \eqref{Bsystem} decouples as follows:
\begin{equation}
	\begin{cases}
    \begin{aligned}
    \label{Bsystemsolv}
	&{{dp_i}\over{d\phi}}=-{{p_i}\over\gamma p_{\phi}}(\mathcal{K}_{j}+\mathcal{K}_{k})\\
	&{{dc_i}\over{d\phi}}={c_{i}\over{\gamma p_{\phi}}}(\mathcal{K}_{j}+\mathcal{K}_{k})
	\end{aligned}
\end{cases}
\end{equation}
for $i\neq j\neq k$ and $\eqref{Bsystemsolv}$ is made analytically solvable. In particular, Fig.$\,$\ref{ashtekar} shows that the Universe volume $V=\sqrt{p_{1}p_{2}p_{3}}$ follows the classical singular behavior also in the Ashtekar variables.

\section{SEMICLASSICAL POLYMER DYNAMICS OF THE BIANCHI I MODEL IN THE ASHTEKAR VARIABLES\label{sem1}}

In this section we present the original part of this paper by investigating the dynamics of the Bianchi I model in terms of the Ashtekar variables when the polymer paradigm is implemented. We recall that this analysis is expected to provide significantly insight on the behavior of the quantum expectation values in PQM as well in LQC dynamics.

\begin{figure}[h!]
	\centering
	\includegraphics[width=0.8\linewidth]{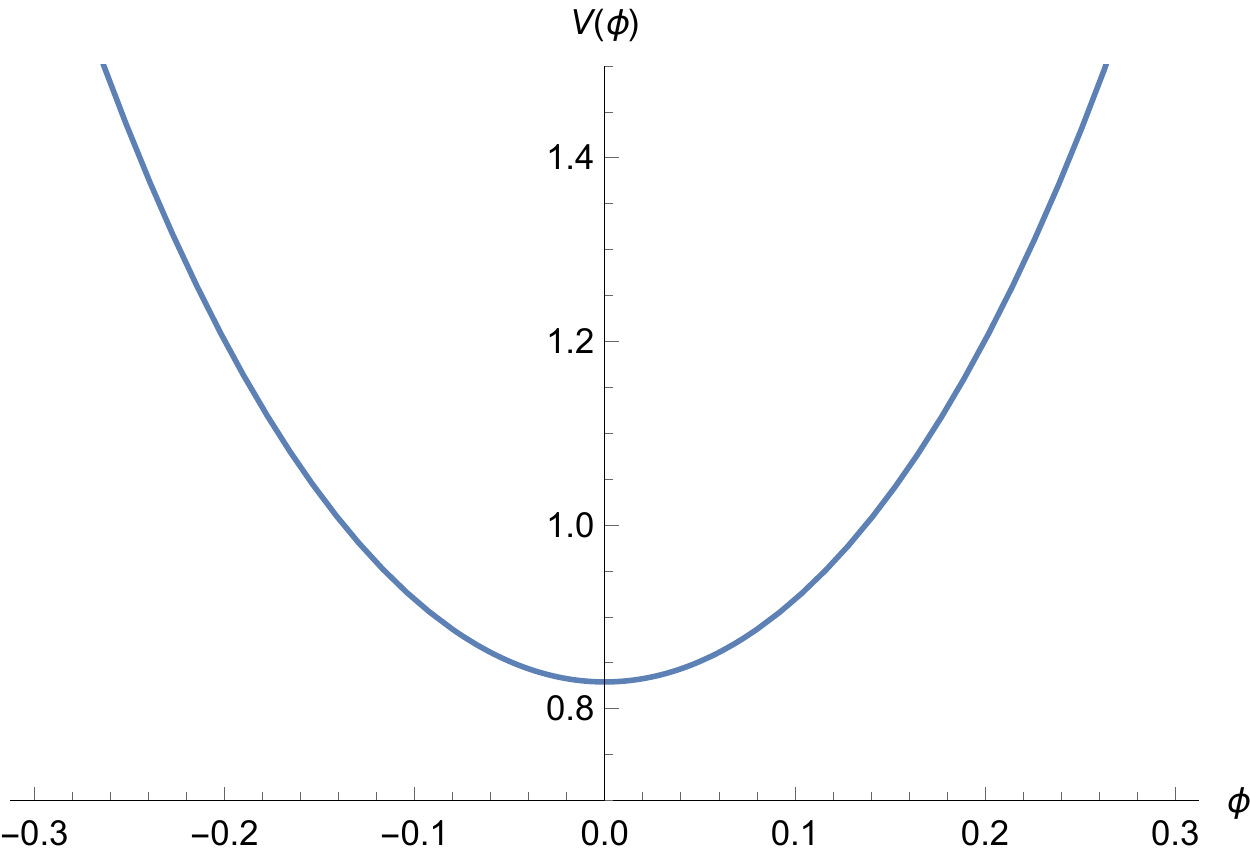}
	\caption{Polymer trajectory of the Universe volume $V=\sqrt{|p_{1}p_{2}p_{3}|}$ in function of time $\phi$: the Big Bounce replaces the classical singularity of the Bianchi I model. In this graph we have set $\kappa=\gamma=1,\mu_1=1/2,\mu_2=1/3,\mu_3=1/4,p_\phi=1/\sqrt{2},\bar{p}_1=1,\bar{p}_2=2$.}
	\label{V(phi)}
\end{figure}

As seen for the FLRW case in Sec. \ref{iso}, we proceed by imposing the polymer substitution for the configurational variables in \eqref{HB}		
\begin{equation}
   c_i\to{1\over\mu_i}sin({\mu_ic_i})\,,
\end{equation}
so the polymer Hamiltonian takes the form
\begin{align}
	\label{eqn:Hpoly}
	\mathcal{H}_{poly}=&-{1\over{\kappa\gamma^2V}}\sum_{i\neq j}{{\sin(\mu_ic_i)p_i\sin(\mu_jc_j)p_j}\over\mu_i\mu_j}+{p^2_{\phi}\over 2V}=0\,,
\end{align}
where $i,j=1,2,3$ and $V=\sqrt{p_1p_2p_3}$. Similarly, $\phi$ represents our internal time, so $N$ is fixed by the gauge
\begin{equation}
	\label{shiftf}
	\dot \phi:=N{{\partial \mathcal{H}_{poly}}\over{\partial p_\phi}}=1\Rightarrow N={{\sqrt{p_1p_2p_3}}\over p_\phi}\,,
\end{equation}
in a way that the equations of motion in the polymer representation are the following:
\begin{equation}
\label{Bsystem}
\begin{cases}
	\begin{aligned}
	\small &{{dp_i}\over{d\phi}}=-{{p_i\cos(\mu_ic_i)}\over\gamma p_{\phi}}\Big[{p_j\over{\mu_j}}\sin(\mu_jc_j)+{p_k\over{\mu_k}}\sin(\mu_kc_k)\Big]\\
	\small &{{dc_i}\over{d\phi}}={\sin(\mu_ic_i)\over{\gamma\mu_i p_{\phi}}}\Big[{p_j\over{\mu_j}}\sin(\mu_jc_j)+{p_k\over{\mu_k}}\sin(\mu_kc_k)\Big]
    \end{aligned}
\end{cases}
\end{equation}
for $i,j,k=1,2,3$, $i\neq j\neq k$. It is possible to solve this system by establishing the initial conditions on the variables $(c_{i},p_{i})$ that must satisfy the Hamiltonian constraint \eqref{eqn:Hpoly}. In this respect, we make the choice\footnote{The considered initial conditions on $c_i$ restrict the solutions to those which are synchronized, due to the need of having a simultaneous Bounce in all the three directional scale factors.}
\begin{equation}
	\label{condin}
	\begin{aligned}
	&c_i(0)={\pi\over 2 \mu_i}\,,\quad p_1(0)=\bar{p}_1\,,\quad p_2(0)=\bar{p}_2\,,\\
	&p_3(0)=\bar{p}_3=\frac{(p_\phi^2\kappa\gamma^2\mu_1\mu_2-2\bar{p_1}\bar{p}_2)\mu_3}{2(\mu_2\bar{p}_1+\mu_1\bar{p}_2)}\,.
	\end{aligned}
\end{equation}
Moreover, it can be easily seen that the momentum conjugate to the scalar field is a first integral, since the variable $\phi$ is cyclic in \eqref{eqn:Hpoly}, and other constants of motion can be obtained by combining the Hamilton equations \eqref{Bsystem}. So, in analogy with \eqref{eqn:integraliprimi}
we get
\begin{equation}
	{p_{i}\sin(\mu_i c_i)\over{\mu_i}}=\mathcal{K}_{i}\,,\hspace{0.5cm}p_{\phi}=\mathcal{K}_{\phi}\,,
\end{equation}
where the considered values of $\mathcal{K}_{i}$ and $\mathcal{K}_{\phi}$ depends on the initial conditions \eqref{condin}. As already mentioned, identifying these first integrals makes possible to transform the six-equations system showed in \eqref{Bsystem} in the three closed systems 
\begin{equation}
\label{Bsystemdecoupled}
\begin{cases}
\begin{aligned}
	\small &{{dp_i}\over{d\phi}}=-{{p_i\cos(\mu_i c_i)}\over\gamma p_{\phi}}\Big[\mathcal{K}_j+\mathcal{K}_k\Big]\\
	\small &{{dc_i}\over{d\phi}}={\sin(\mu_i c_i)\over{\gamma\mu_i p_{\phi}}}\Big[\mathcal{K}_j+\mathcal{K}_k\Big]
\end{aligned}
\end{cases}
\end{equation}
Thanks to this procedure, the equations of motion can be solved analytically, leading to the following solutions ($i\neq j, i,j=1,2$):
\begin{equation}
	\label{sysp}
	\begin{aligned}
		&c_i(\phi)=\frac{2}{\mu_i}\text{arccot}\big[\text{exp}\big(-\frac{\bar{p}_3/\mu_3+\bar{p}_j/\mu_j}{\gamma p_\phi}\phi\big)\big]\,,\\ 
		&p_i(\phi)=\bar{p}_i\cosh\Big[\frac{\bar{p}_3/\mu_3+\bar{p}_j/\mu_j}{\gamma p_\phi}\phi\Big]\,;\\
		&c_3(\phi)=\frac{2}{\mu_3}\text{arccot}\big[\text{exp}\big(-\frac{\bar{p}_1/\mu_1+\bar{p}_2/\mu_2}{\gamma p_\phi}\phi\big)\big]\,,\\
		&p_3(\phi)=\bar{p}_3\cosh\Big[\frac{(\bar{p}_1/\mu_1+\bar{p}_2/\mu_2)}{\gamma p_\phi}\phi\Big]\,.\\
	\end{aligned}
\end{equation}
Using \eqref{sysp}, we can obtain the Universe volume behavior $V(	\phi)=\sqrt{|p_1(\phi)p_2(\phi)p_3(\phi)|}$ that is shown in Fig.$\,$\ref{V(phi)}. The resulting trajectory highlights that a semiclassical Big Bounce replaces the classical Big Bang thanks to the regularizing polymer effects, which are expected to become dominant near the Planckian region.

\section{SEMICLASSICAL POLYMER DYNAMICS OF THE BIANCHI I MODEL IN THE VOLUME-LIKE VARIABLES \label{sem2}}

In this section we study the dynamics of the Bianchi I Universe for a new choice of variables, in complete analogy with the analysis performed for the FLRW model in Sec. \ref{FRWV}. More specifically, the anisotropic character of the Bianchi I model leads to the possibility of taking into account two different sets of volume-like variables, that coincide in the case of the isotropic model. Then, we will compare the obtained results.

\subsection{Analysis in the anisotropic volume-like variables: $(V_{1},V_{2},V_{3})$}

Firstly, we consider as a set of volume-like variables three equivalent generalized coordinates which coalesce to the proper volume in the isotropic limit only (see \cite{szulc}):
\begin{equation}
	V_{i}=\sign(p_{i})|p_{i}|^{3\over2}\,,\hspace{0.5cm}\beta_{i}={2c_{i}\over3\sqrt{|p_{i}|}}\,,
\end{equation}
where $\beta_i$ for $i=1,2,3$ are the conjugate momenta and the new symplectic structure for the system is characterized by the conserved Poisson brackets $\{\beta_{i},V_{j}\}=\kappa\gamma\delta_{ij}$\,. In this case we are not promoting one of the configurational variables to represent the Universe volume. On the contrary, we are imposing that the three independent coordinates are isomorphic to the isotropic volume for each direction, so that $V=|V_{1}V_{2}V_{3}|^{1\over3}$.

The Hamiltonian constraint for this framework in the semiclassical polymer representation is obtained by using the polymer substitution for the momenta $\beta_i$ after that the canonical transformation on \eqref{HB} has been performed, and it reads as
\begin{equation}
	\label{Hv}
	\mathcal{H}_{poly}=-{9\over{4\kappa\gamma^{2}V}}\sum_{i\neq j}{V_{i}\sin(\mu_{i}\beta_{i})V_{j}\sin(\mu_{j}\beta_{j})\over\mu_{i}\mu_{j}}+{p_{\phi}^{2}\over 2 V}=0\,,
\end{equation}
where $i,j=1,2,3$.
We impose $N=\frac{V}{p_\phi}$ due to the choice of $\phi$ as relational time, so the Hamilton equations describing the dynamics are
\begin{equation}
	\label{sysV}
	\begin{cases}
		\begin{aligned}
			&\frac{dV_i}{d\phi}=-{{9V_i\cos(\mu_i\beta_i)}\over 4\gamma p_{\phi}}\bigg[{V_j\over{\mu_j}}\sin(\mu_j\beta_j)+{V_k\over{\mu_k}}\sin(\mu_k\beta_k)\bigg]\\
			&\frac{d\beta_i}{d\phi}={9\sin(\mu_i\beta_i)\over{4\gamma\mu_ip_{\phi}}}\bigg[{V_j\over{\mu_j}}\sin(\mu_j\beta_j)+{V_k\over{\mu_k}}\sin(\mu_k\beta_k)\bigg]
		\end{aligned}
	\end{cases}
\end{equation}
for $i\neq j\neq k$. In analogy with the previous treatment, we can identity the following constants of motion:
\begin{equation}
	{V_{i}\sin(\mu_i\beta_i)\over{\mu_i}}=\mathcal{K}_{i}\,,\hspace{0.5cm}p_{\phi}=\mathcal{K}_{\phi}\,,
\end{equation}
that decouple the system \eqref{sysV}  along the three directions. By taking general initial conditions according to \eqref{Hv}, the analytical solutions for the anisotropic volume coordinates read as:
\begin{equation}
	\label{aniV}
	\begin{aligned}
		&V_1(\phi)=\bar{V}_1\cosh\Big[\frac{9(\bar{V}_3/\mu_3+\bar{V}_2/\mu_2)}{4\gamma p_\phi}\phi\Big]\,,\\
		&V_2(\phi)=\bar{V}_2\cosh\Big[\frac{9(\bar{V}_3/\mu_3+\bar{V}_2/\mu_2)}{4\gamma p_\phi}\phi\Big]\,,\\
		&V_3(\phi)=\bar{V}_3\cosh\Big[\frac{9(\bar{V}_1/\mu_1+\bar{V}_2/\mu_2)}{4\gamma p_\phi}\phi\Big]\,,\\
	\end{aligned}
\end{equation}
where 
\begin{equation}
	\label{condinVi}
	\begin{aligned}
		&\beta_i(0)={\pi\over 2 \mu_i}\,,\quad V_1(0)=\bar{V}_1\,,\quad V_2(0)=\bar{V}_2\,,\\
		&V_3(0)=\bar{V}_3=\frac{(2p_\phi^2\kappa\gamma^2\mu_1\mu_2-9\bar{V}_1\bar{V}_2)\mu_3}{9\big(\mu_2\bar{V}_1+\mu_1\bar{V}_2\big)}\,.
	\end{aligned}
\end{equation}

\begin{figure}[htbp]
	\centering
	\includegraphics[width=0.8\linewidth]{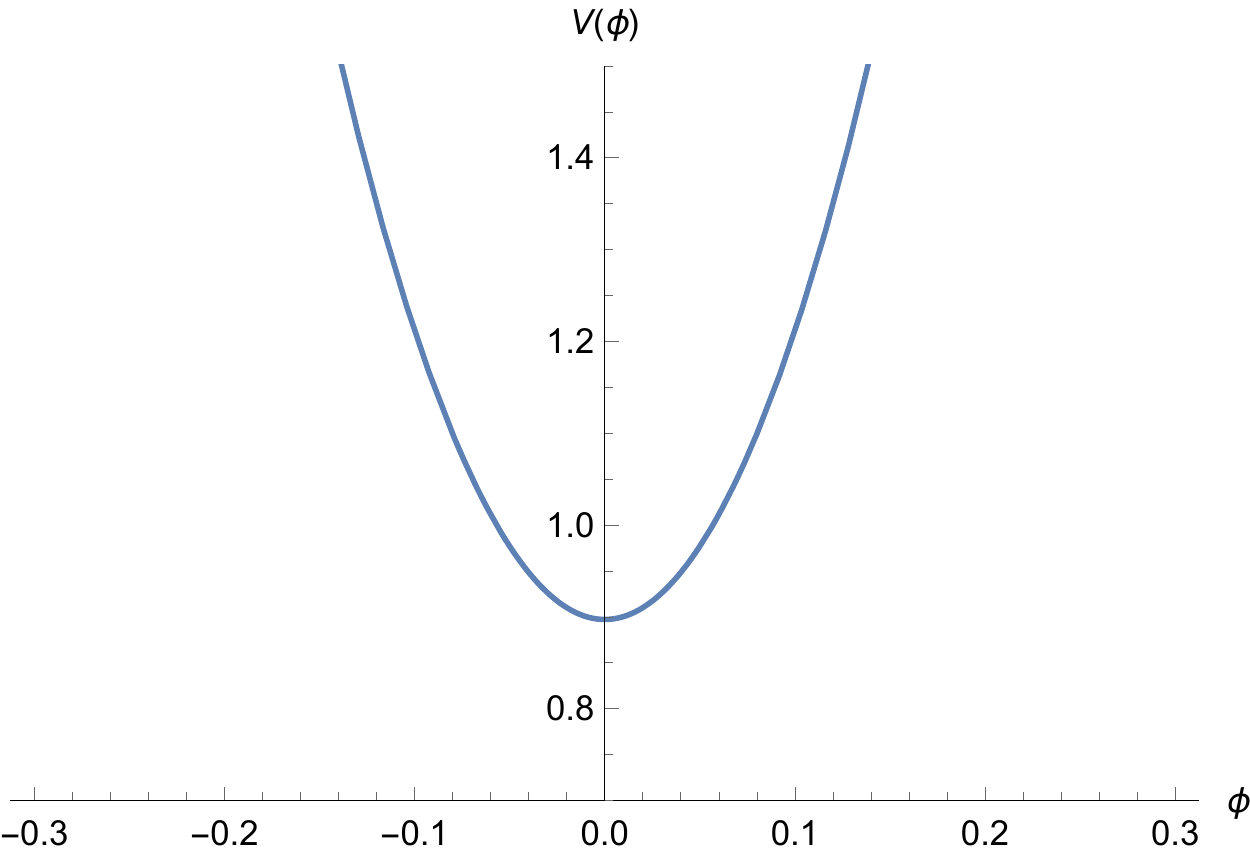}
	\caption{Semiclassical polymer trajectory of the Universe volume $V=|V_1V_2V_3|^{1/3}$ as function of $\phi$. In this graph we have set $\kappa=\gamma=1,\mu_1=1/2,\mu_2=1/3,\mu_3=1/4,p_\phi=1/\sqrt{2},\bar{V}_1=1,\bar{V}_2=2$.}
	\label{Vv}
\end{figure}
By combining the solutions \eqref{aniV} we can find the Universe volume behavior in function of $\phi$ as $V(\phi)=|V_1(\phi)V_2(\phi)V_3(\phi)|^{1/3}$, that is characterized by the emergence of a Big Bounce as shown in Fig.$\,$\ref{Vv}.

\subsection{Analysis in the volume variables: $(v,\lambda_{1},\lambda_{2})$}
\label{SecV}
The proper set of volume variables is defined as
\begin{equation}
	\label{pv}	
	\begin{aligned}
	&\lambda_{1,2}=\sign(p_{1,2})\sqrt{|p_{1,2}|}\,, \hspace{0.4cm} v=\sqrt{|p_1p_2p_3|}\,,\\
	&\eta_{1,2}=\frac{2p_{1,2}c_{1,2}-2p_3c_3}{\sqrt{|p_{1,2}|}}\,,\quad\eta_3=2\sqrt{\bigg|\frac{p_3}{p_1p_2}\bigg|}c_3\,.
	\end{aligned}
\end{equation}
The Poisson brackets are conserved $(\{\eta_i,\lambda_j\}=\kappa\gamma\delta_{ij}$ and $\{\eta_i,\eta_j\}=\{\lambda_i,\lambda_j\}=0$, with $\lambda_3=v$ and $i,j=1,2,3$) and the semiclassical polymer Hamiltonian takes the form
\begin{equation}
	\begin{aligned}
	\label{HV}
	&\mathcal{H}_{poly}=-{1\over{4\kappa\gamma^{2}v}}\Big(\sum_{i=1,2}2{\lambda_{i}\sin(\mu_{i}\eta_{i})v\sin(\mu_{3}\eta_{3})\over\mu_{i}\mu_{3}}+\\&+{\lambda_{1}\sin(\mu_{1}\eta_{1})\lambda_{2}\sin(\mu_{2}\eta_{2})\over\mu_{1}\mu_{2}}+3v^2\frac{\sin^2(\mu_3\eta_3)}{\mu_3^2}\Big)+{p_{\phi}^{2}\over 2 v}=0\,,
	\end{aligned}
\end{equation}
where the canonical transformation has been performed before the implementation of the semiclassical polymer paradigm, as in the previous subsection. We note that the Hamiltonian constraint \eqref{HV} has not the same expression of the one proposed in \cite{ashtekar2009}, since the same variables as $(v,\lambda_1,\lambda_2)$ are used but different ones are polymerized (namely $\eta_1$ and $\eta_2$ have not the same meaning).

Analogously, we derive the Hamilton equations for the couple of variables ($v,\eta_{3}$)
\begin{equation}
		\label{vphi}
		\begin{cases}
				\begin{aligned}
	&\frac{dv}{d\phi}=-{{v\cos(\mu_3\eta_3)}\over 4\gamma p_{\phi}}\bigg[2\sum_{i=1,2}{\lambda_i\over{\mu_i}}\sin(\mu_i\eta_i)+6{v\over{\mu_3}}\sin(\mu_3\eta_3)\bigg]\\
	&\frac{d\eta_3}{d\phi}={\sin(\mu_3\eta_3)\over{4\gamma\mu_3p_{\phi}}}\bigg[2\sum_{i=1,2}{\lambda_i\over{\mu_i}}\sin(\mu_i\eta_i)+6{v\over{\mu_3}}\sin(\mu_3\eta_3)\bigg]
	\end{aligned}
\end{cases}
\end{equation}
and also for the conjugate variables $(\lambda_1,\eta_1)$, $(\lambda_2,\eta_2)$
\begin{equation}
	\begin{cases}
	\begin{aligned}
	&\frac{d\lambda_i}{d\phi}=-{{\lambda_i\cos(\mu_i\eta_i)}\over 4\gamma p_{\phi}}\bigg[2{v\over{\mu_3}}\sin(\mu_3\eta_3)+{\lambda_{j}\over{\mu_j}}\sin(\mu_j\eta_j)\bigg]\\
	&\frac{d\eta_i}{d\phi}={\sin(\mu_i\eta_i)\over{4\gamma\mu_ip_{\phi}}}\bigg[2{v\over{\mu_3}}\sin(\mu_3\eta_3)+{\lambda_{j}\over{\mu_j}}\sin(\mu_j\eta_j)\bigg]
	\end{aligned}
\end{cases}
\end{equation}
where we have used
$N=\frac{v}{p_\phi}$
in order to derive the dynamics of the model in function of the relational time $\phi$.

Once fixed the initial conditions on the variables $(\lambda_1,\eta_1)$, $(\lambda_2,\eta_2)$, $(v,\eta_3)$ according to \eqref{HV}, we can solve this system analytically since the three-dimensional motion is decoupled in three one-dimensional trajectories thanks to the use of analogous constants of motion:
\begin{equation}
	{\lambda_{1,2}\sin(\mu_{1,2}\eta_{1,2})\over{\mu_{1,2}}}=\mathcal{K}_{1,2}\,,\hspace{0.3cm}{v\sin(\mu_3\eta_3)\over{\mu_3}}=\mathcal{K}_{3}\,,\hspace{0.3cm}p_{\phi}=\mathcal{K}_{\phi}\,.
\end{equation}		
In this case, we fix the constants of motion as follows:
\begin{equation}
	\begin{aligned}
		\label{cond}
		&\mathcal{K}_1=\bar{\mathcal{K}}\sqrt{\frac{2\kappa\gamma^2p_\phi^2+\mathcal{K}^2}{(\bar{\mathcal{K}}+1)(\bar{\mathcal{K}}+3)}}+\mathcal{K}\,,\\
		&\mathcal{K}_2=\bar{\mathcal{K}}\sqrt{\frac{2\kappa\gamma^2p_\phi^2+\mathcal{K}^2}{(\bar{\mathcal{K}}+1)(\bar{\mathcal{K}}+3)}}-\mathcal{K}\,,\\
		&\mathcal{K}_3=\sqrt{\frac{2\kappa\gamma^2p_\phi^2+\mathcal{K}^2}{(\bar{\mathcal{K}}+1)(\bar{\mathcal{K}}+3)}}\,,
	\end{aligned}
\end{equation}
where $\bar{\mathcal{K}}\,,\mathcal{K}$ are two free parameters. As we will see below, a convenient form for the polymer-modified Friedmann equation can be obtained thanks to this particular choice for the constants of motion; however, the physical properties that will be derived still have a general meaning, since they are valid for all the values assigned to the constants $\bar{\mathcal{K}}\,,\mathcal{K}$ (with $\bar{\mathcal{K}}\neq-1,-3$). In particular, in the following it will be highlighted the existence of a non-trivial solution to the equation $H^2=0$, that identifies the expression of the critical energy density for which the scale factor velocity becomes null. More precisely, this information allows to identify also the anisotropy contribution to the total critical energy density added to the standard one associated to the matter fields. This way, it is possible to rigorously analyze the physical properties of the critical point, whose presence is due to the polymer cut-off effects.

Regarding the Bianchi I model, the standard Friedmann equation reads as (see \cite{G})
\begin{equation}
	H^{2}={\kappa\over 3}(\rho+\rho_{aniso})\,,
	\label{Bfrie}
\end{equation}  
where the additional term $\rho_{aniso}$ accounts for the contribution of the anisotropic gravitational degrees of freedom to the total energy density. In what follows, we want to verify how the semiclassical polymer approach modifies equation \eqref{Bfrie} and, in addition, if the total critical energy density derived from the modified Friedmann equation has universal properties when the Universe volume itself is considered as a configurational variable.

\begin{figure}[htbp]
	\centering
	\includegraphics[width=0.8\linewidth]{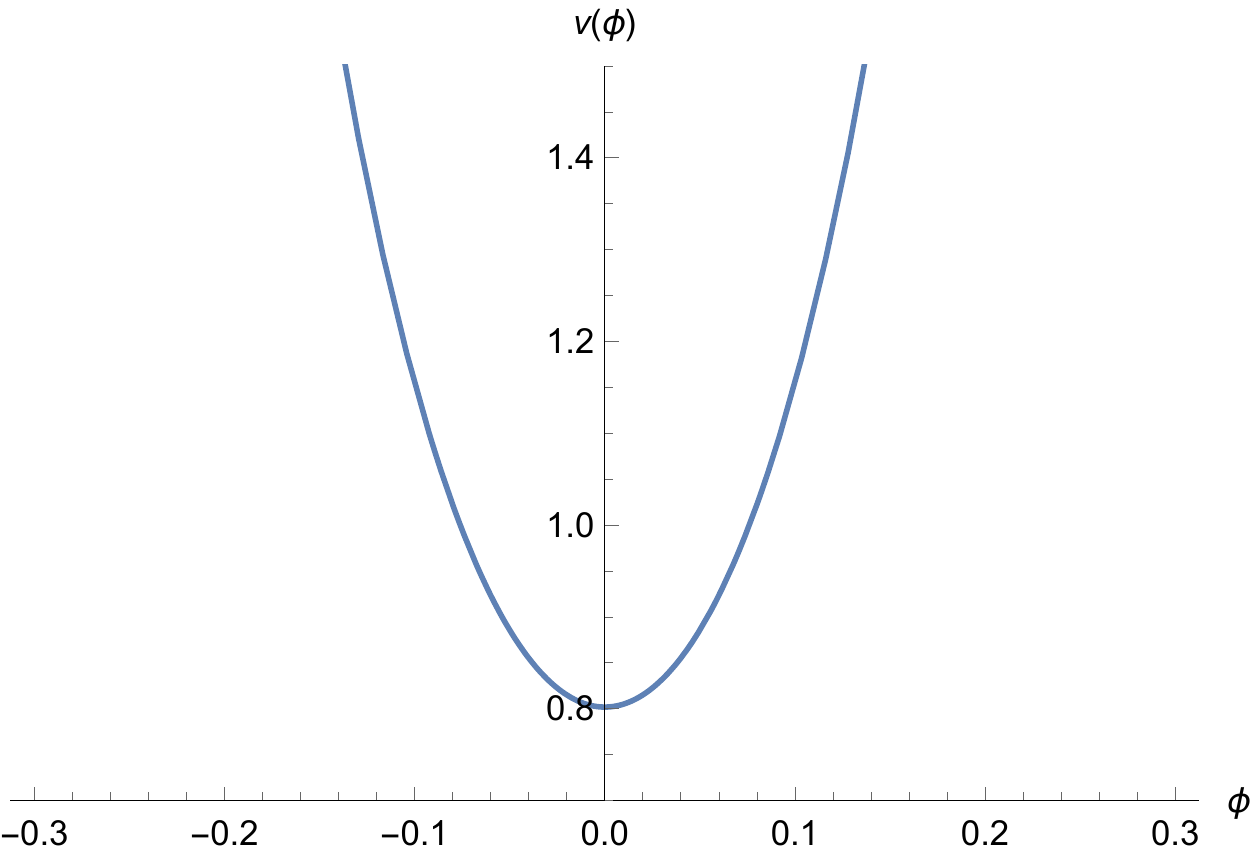}
	\caption{Semiclassical polymer trajectory of the Universe volume $v(\phi)$. In this graph we have set $\gamma=1,\mu_3=1/4,p_\phi=1/\sqrt{2},\mathcal{K}_3=3,\bar{\mathcal{K}}=1/2$.}
	\label{vol}
\end{figure}

In this set of volume variables, the Hubble parameter can be written as
\begin{equation}
	\begin{aligned}
		H^2=\Big(\frac{1}{3v}\frac{dv}{dt}\Big)^2&=\frac{(\mathcal{K}_1+\mathcal{K}_2+3\mathcal{K}_3)^2}{36\gamma^2v^2}\cos^2(\mu_3\eta_3)=\\
		&=\frac{(\mathcal{K}_1+\mathcal{K}_2+3\mathcal{K}_3)^2}{36\gamma^2v^2}[1-\sin^2(\mu_3\eta_3)]=\\&=\frac{(\mathcal{K}_1+\mathcal{K}_2+3\mathcal{K}_3)^2}{36\gamma^2v^2}(1-\frac{\mu_3^2}{v^2}\mathcal{K}_3^2)
		\label{BF}
	\end{aligned}
\end{equation}
where we restored the synchronous time-gauge $N=1$ in the equation for the volume written in \eqref{vphi}. Now, if we substitute the conditions expressed above in \eqref{cond} we obtain
\begin{equation}
	\label{FRIEDMANN}
	\begin{aligned}
	&H^2=\frac{\kappa}{18}\frac{f_1(\bar{\mathcal{K}})^2}{f_2(\bar{\mathcal{K}})}\frac{p_\phi^2+p_{aniso}^2}{v^2}\Big[1-\frac{4\kappa\gamma^2\mu_3^2}{f_2(\bar{\mathcal{K}})}\Big(\frac{p_\phi^2+p_{aniso}^2}{2v^2}\Big)\Big]\,,\\
	&p_{aniso}={\mathcal{K}\over{\sqrt{2\kappa\gamma^2}}}
	\end{aligned}
\end{equation}
where  $f_1(\bar{\mathcal{K}})=2\bar{\mathcal{K}}+3\,,f_2(\bar{\mathcal{K}})=(\bar{\mathcal{K}}+1)(\bar{\mathcal{K}}+3)\,.$
This expression represents the polymer-modified Friedmann equation for the Bianchi I model in the proper volume variables. In particular, the additional term $p_{aniso}^2/2v^2$ reasonably mimics the anisotropic contribution $\rho^{aniso}$, so that we can compute $\rho^{tot}_{crit}$ as
\begin{equation}
	\label{tced}
	\rho^{tot}_{crit}=\rho^\phi_{crit}+\rho^{aniso}_{crit},
\end{equation}
where the term regarding the matter scalar field takes the usual expression $p_\phi^2/(2v^2)$.
Moreover, from \eqref{FRIEDMANN} the total critical energy density results to be
\begin{equation}
	\rho^{tot}_{crit}=\frac{f_2(\bar{\mathcal{K}})}{4\kappa\gamma^2\mu_3^2}\,.
	\label{tcedv}
\end{equation}
When the initial conditions on the motion satisfy \eqref{cond}, the solution for the Universe volume $v(\phi)$ is
\begin{equation}
	v(\phi)=\mu_3|\mathcal{K}_3|\,\cosh\bigg(\frac{f_1(\bar{\mathcal{K}})\mathcal{K}_3}{2\gamma p_\phi}\,\phi\bigg)
	\label{vphii}
\end{equation}
and it clearly resembles a bouncing behavior as shown in Fig. \ref{vol}.

Now we can take the expression of the total energy density from \eqref{FRIEDMANN} and use the law of motion for the Universe volume to compute its critical value:
\begin{equation}
	\begin{aligned}
	\rho^{tot}_{crit}&=\frac{p_\phi^2}{2v(\phi)|_{\phi=0}^2}+\frac{p_{aniso}^2}{2v(\phi)|_{\phi=0}^2}=\\&=\frac{f_2(\bar{\mathcal{K}})p_\phi^2}{2\mu_3^2(2\kappa\gamma^2p_\phi^2+\mathcal{K}^2)}+\frac{f_2(\bar{\mathcal{K}})\mathcal{K}^2}{4\kappa\gamma^2\mu_3^2(2\kappa\gamma^2p_\phi^2+\mathcal{K}^2)}\,=\\&=\frac{f_2(\bar{\mathcal{K}})}{4\kappa\gamma^2\mu_3^2}\,,
	\end{aligned}
\end{equation}
that is equal to \eqref{tcedv} as expected. Therefore, this analysis shows that taking the Universe volume itself as a configurational variable makes the Big Bounce acquire universal physical properties, in agreement with the behavior obtained for the set $(v,\eta)$ in the FLRW model (see Sec. \ref{FRWV}). 

\section{Equivalence feature of different sets of variables}\label{discussion}
In Secc. \ref{sem1}-\ref{sem2} we performed the semiclassical polymer analysis of the Bianchi I model with a massless scalar field in three sets of variables, i.e. the Ashtekar one and two different volume-like ones. The bouncing behavior over time $\phi$ for the Universe volume has been outlined for all the three phase space configurations, but the universal properties of the critical point have been shown only in the proper set of volume variables (i.e. when the Universe volume itself is one of the three configurational variables), thanks to the derivation of the polymer-modified Friedmann equation \eqref{FRIEDMANN}. Obviously, the three sets of conjugate variables are canonically related at a classical level, but this is a necessary but not sufficient condition for obtaining equivalent dynamics in the polymer formulation (see \cite{crino,EBianchiIX,Ant,Review} for not equivalent polymer cosmologies). In other words, performing the canonical transformation and doing the polymer substitution do not commute, as shown in \cite{EFG}. Indeed, it can be verified that both the Hamiltonian constraint and the Hamilton's equations written in the three sets of variables cannot be mapped in the semiclassical polymer framework (see Secc. \ref{sem1}-\ref{sem2}).

The aim of this section is to check whether it is possible to recover the equivalence between the polymer semiclassical dynamics analyzed in Secc. \ref{sem1}-\ref{sem2} by generalizing the approach used in \cite{EFG}. We start by considering the relation between the Ashtekar variables and the anisotropic volume-like ones (for $i=1,2,3$), i.e.
\begin{equation}
	|p_i|=V_i^{2/3}\,, \hspace{0.5cm} c_i=\frac{3}{2}V_i^{1/3}\beta_i
\end{equation}
that ensures the canonical equivalence of the two frameworks at a classical level. In order to preserve the formal invariance of the Poisson brackets also in the polymer formulation ($\beta_i\rightarrow\frac{1}{\mu_i}\sin(\mu_i\beta_i)$ and $c_i\rightarrow\frac{1}{\mu_i'}\sin(\mu_i'c)$), i.e. 
\begin{equation}
\{\beta_i,V_i\}=\kappa\gamma\sqrt{1-(\mu_i\beta_i)^2}=\kappa\gamma\sqrt{1-(\mu_i'c_i)^2}=\{c_i,p_i\}\,,
\end{equation}
we impose the conditions $\mu_i\beta_i=\mu'_ic_i$. So, we obtain that the polymer parameters related to the Ashtekar connections depend on the configurational variables as 
\begin{equation}
\mu'_i=\frac{2\mu_{i}}{3\sqrt{|p_i|}}
\label{muprimo2}
\end{equation}
where $\mu_1,\mu_2,\mu_3$ are constant. So, by using \eqref{muprimo2} the Hamiltonian in the Ashtekar variables \eqref{eqn:Hpoly} and that one in the anisotropic volume-like ones \eqref{Hv} are mapped also after that the polymer substitution has been implemented, i.e. supposing that the polymer parameters are not invariant under the change of variables makes commutative to write the Hamiltonian in the new set of variables and to implement the semiclassical polymer substitution. This way, we have demonstrated the semiclassical equivalence between the dynamics in the Ashtekar variables and that in the anisotropic volume-like ones, once considered the polymer parameters in the Ashtekar variables depending on the configurational coordinates.
Clearly, also the equations of motion are mapped in the two frameworks and the same dynamical properties are obtained by using the new or the old Hamiltonian. Indeed, we have (for $i,j,k=1,2,3$, $i\neq j\neq k$)
\begin{equation}
\begin{cases}
\begin{aligned}
{{dp_i}\over{d\phi}}&=\frac{2\sign(V_i)\dot{V}_i}{3V_i^{1/3}}=\kappa\gamma\frac{V}{p_\phi}\frac{\partial\mathcal{H}_{poly}(c_i,p_i)}{\partial c_i}=\\&=-{{p_i\cos(\mu'_ic_i)}\over\gamma p_{\phi}}\Big[{p_j\over{\mu'_j}}\sin(\mu'_jc_j)+{p_k\over{\mu'_k}}\sin(\mu'_kc_k)\Big]\\
{{dc_i}\over{d\phi}}&=\frac{3}{2}\Big(V_i^{1/3}\dot{\beta}_i+\beta_i\frac{\dot{V}_i}{V_i^{2/3}}\Big)=-\kappa\gamma\frac{V}{p_\phi}\frac{\partial\mathcal{H}_{poly}(c_i,p_i)}{\partial p_i}=\\&=\frac{3}{2}\Big({\sin(\mu'_ic_i)\over{\gamma\mu'_i p_{\phi}}}-{c_i\cos(\mu'_ic_i)\over{\gamma p_{\phi}}}\Big)\Big[{p_j\over{\mu'_j}}\sin(\mu'_jc_j)+{p_k\over{\mu'_k}}\sin(\mu'_kc_k)\Big]
\end{aligned}
\end{cases}
\end{equation}
where we notice that the equation of motion for $p_i$ is formally equivalent to that one in \eqref{Bsystem} (but the solutions will be different due to the new definition \eqref{muprimo2} of $\mu_i'$), while the equation of motion for $c_i$ is different from that one in \eqref{Bsystem} because of the dependence of the polymer parameter on $p_i$ (see \eqref{muprimo2}). So, the evolution of the Universe volume is equivalent in the two pictures and the Bounce acquire the same physical properties of that set in which the polymer parameters are constant. We remark that this equivalence remains valid only in the semiclassical formulation of the polymer paradigm, since the promotion of the polymer momentum to a quantum operator will be a subtle issue when the polymer parameter depends explicitly on the coordinate.

Now, we could proceed in an analogous way to show the equivalence between the anisotropic volume-like variables and the proper volume ones, so that also the Ashtekar formulation and the volume one would be dynamically related. Unfortunately, as we can see from \eqref{pv} the canonical transformation from the Ashtekar set to the proper volume one combines the directions in the momenta space, with the consequence that $\eta_{1,2}$ is not simply proportional to $c_{1,2}$ by means of a function only of the coordinates. This feature prevents the implementation of the formalism traced above, since the polymer theory is based on one-dimensional lattices by construction. So, at the level of the polymer paradigm it is not possible to investigate the equivalence between the proper volume set and the Ashtekar one (or the anistropic volume-like one), not even in their semiclassical formulations.

By concluding, this analysis demonstrates that imposing a cut-off on configurational variables that have the geometrical dimensions of areas and volumes is sufficient to reproduce a bouncing behavior for the Bianchi I model, but the universal bound on the total critical energy density is ensured only when the proper Universe volume is defined on a polymer lattice with constant spacing. Moreover, by admitting a dependence of the polymer parameters on the configurational coordinates it has been demonstrated the dynamical equivalence between the Ashtekar formulation and the anisotropic volume-like one, due to the fact that the canonical change of variables from the areas $p_i$ to the volume-like variables $V_i$ does not mix the momenta directions, as in the case of the proper volume set. Nevertheless, the development of a more general polymer formulation would be a good instrument to investigate the equivalence between the proper volume variables and the Ashtekar ones, i.e. the original variables of the LQG theory. This way, differently from the picture obtained when constant polymer parameters are implemented, it would be guaranteed the presence of a universal Bounce also in the Ashtekar set at least at a semiclassical level and so inferred the equivalence between the two LQC schemes, at least at an effective level.
\section{QUANTUM ANALYSIS\label{quant}} 	
In order to perform the polymer quantization of the Bianchi I model in the Ashtekar variables, firstly we have to face the question of defining a time variable that is suitable to the description of quantum dynamics (regarding the problem of time in quantum gravity, see \cite{Kuchar,Isham1993}). Despite the Universe volume is often adopted as a good time variable \cite{G,PC,2008}, we observe that it is not a suitable clock for the cosmological dynamics when a bouncing cosmology is analyzed. Indeed, across the Bounce the collapse of the Universe is followed by a re-expansion and so the Universe volume violates the basic prescription of being a monotonic variable \cite{Isham1993,CQG}. Here, the problem of time is addressed by selecting the scalar field (to be regarded as the kinetic component of the primordial inflaton \cite{2002,2010}) as a relational time in the sense defined in \cite{Rovelli}. This same procedure is considered as the most natural in all the relevant investigations of the Big Bounce in quantum cosmology \cite{Review,Ashtekar2011}. On the other hand, in \cite{Bquantum} the loop quantization of the homogeneous cosmologies is implemented, but a diagonal component of the triads is chosen as internal time when studying the dynamics of the Bianchi I model. The argument used is that the behavior of the triad components remains monotonic in the classical regime and that the absence of the initial singularity is only related to the possibility of uniquely well-defining the wave function throughout the minisuperspace. It is worth noting that in \cite{Bquantum} only the expanding branch of the evolution is considered and the analysis is limited to a full quantum approach, whereas the semiclassical dynamics is not addressed. On the contrary, here it is performed a canonical quantization of the polymer semiclassical constraints and the semiclassical dynamics (in which the inverse triad effects are neglected) shows clearly that the only monotonic variable is the scalar field, since a bouncing dynamics emerges.

Once fixed the natural clock for the system, we have to face a more subtle question that concerns the morphology of the quantum equation we have to employ in describing the quantum features of the Bianchi I model. In the spirit of the Dirac prescription of quantizing a constrained system \cite{Dirac},
we are lead to write down the following Wheeler-DeWitt equation for the Universe wave function
\begin{equation}
\label{eqn:WDW}
\Big[-\partial_{\phi}^2+\frac{2}{\kappa\gamma^2}\Big( \partial_{x_{1}}\partial_{x_{2}}+\partial_{x_{1}}\partial_{x_{3}}+ \partial_{x_{2}}\partial_{x_{3}}\Big)\Big]\Psi=0\,,
\end{equation}
where $\hat{p}_\phi=-i\partial_\phi$, $\hat{p}_i=-i\partial_{c_i}$ and it has been used the substitution
\begin{equation}
	\label{xiS}
	x_i=\ln\bigg[{\tan\bigg({\frac{\mu_i c_i}{2}}}\bigg)\bigg]+\bar{x}_i\,.
\end{equation}
The conserved probability density is
\begin{equation}
	\mathcal{J}_0=i(\Psi^*\partial_\phi\Psi-\Psi\partial_\phi\Psi^*)\,
\end{equation}
and it is not positive defined, as any Klein-Gordon-like measure. To adopt it in a well-defined way, the frequency separation procedure is required and also can be naturally performed in the absence of a potential term associated to the scalar field (that is reasonably negligible at sufficiently high temperature and/or energy densities of the Universe
\cite{PC,Kolb,1999,2000,2000bis}). However, by comparing the quantum dynamics of quasi-classical localized state with the semiclassical Friedmann dynamics, we are able to outline that the singularity is removed and a cut-off in the collapse naturally arises. So, in order to deal with a localized state, we need to follow the evolution of a suitable wave packet and this requires the superposition of different (infinite in a Gaussian packet) frequencies. Now, the subtle question is in the impossibility to ensure the positive nature of the probability density \cite{rosenstein1985probability} when more than one frequency mode is considered, even if the frequencies have all the same sign. This is a well known result in relativistic quantum mechanics \cite{rosenstein1985probability} and it can be easily shown by considering two frequencies only. Even if this unpleasant effect is not evident when we deal with a Gaussian packet, to solve the superHamiltonian constraint before quantizing the system, i.e. to perform an ADM reduction of the dynamics \cite{ADM,G,PC,CQG} seems a more suitable canonical method of quantization in agreement with the Dirac prescription. Before passing to the technical aspects of this approach, which leads to a Schr\"{o}dinger-like equation for the Universe wave function, we observe that in such a procedure we pay the fixing of a specific temporal gauge, which is often thought as a choice concerning the classical metric morphology. In particular, we assume to deal with a lapse function which depends on time according to its dependence on the phase space variables and by virtue of the classical Hamilton's equations. However, when quantizing the system also this fixed lapse function becomes a quantum operator and it has to act on the wave function of the Universe too. This ambiguity in the interpretation of the time gauge fixing is beyond the scope of the present analysis, but we underline it in order to stress how the construction of a parabolic constraint for the quantization, or equivalently the removal of an hyperbolic one, is sometime a more viable approach but has non-trivial interpretative consequences. Actually, from the point of view of the original Wheeler-DeWitt constraint, we are leaving a pure geometrical approach.
\begin{figure*}
	\begin{minipage}{3.4cm}
		\includegraphics[width=1.5\linewidth]{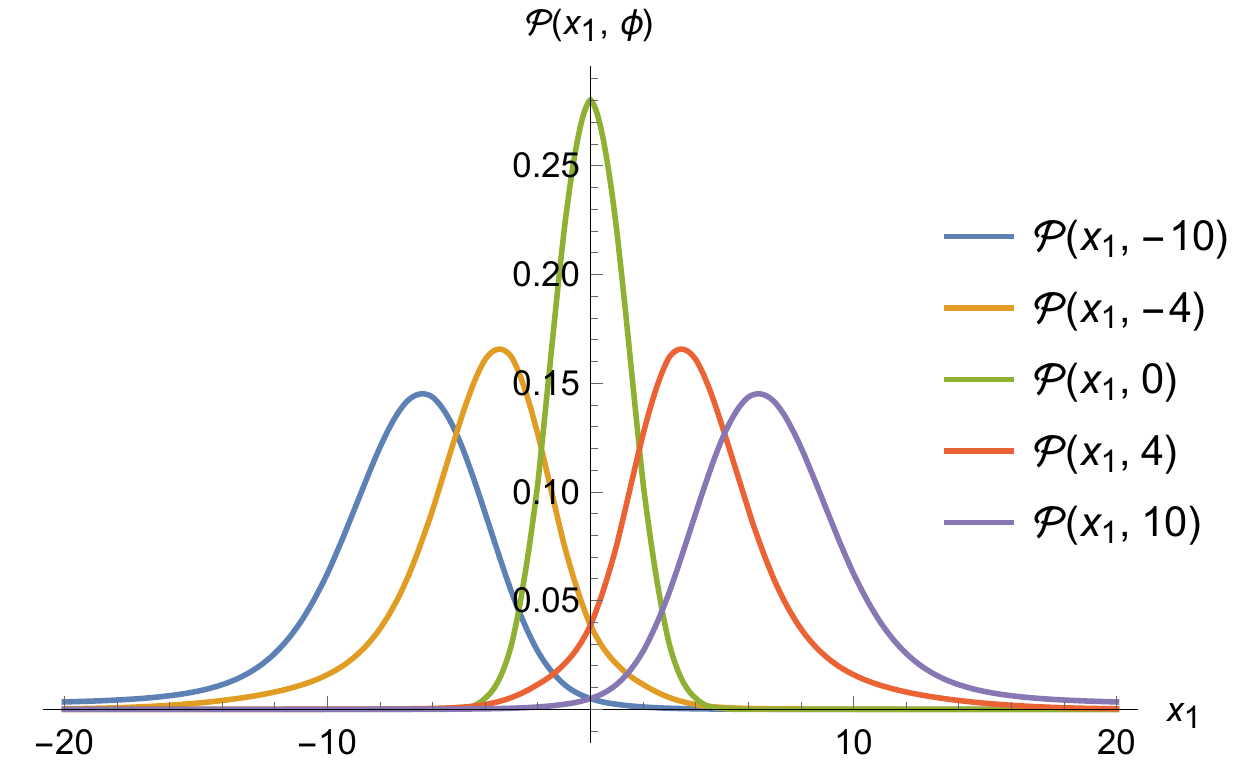}
	\end{minipage}
	\qquad\qquad\qquad
	\begin{minipage}{3.4cm}
		\includegraphics[width=1.5\linewidth]{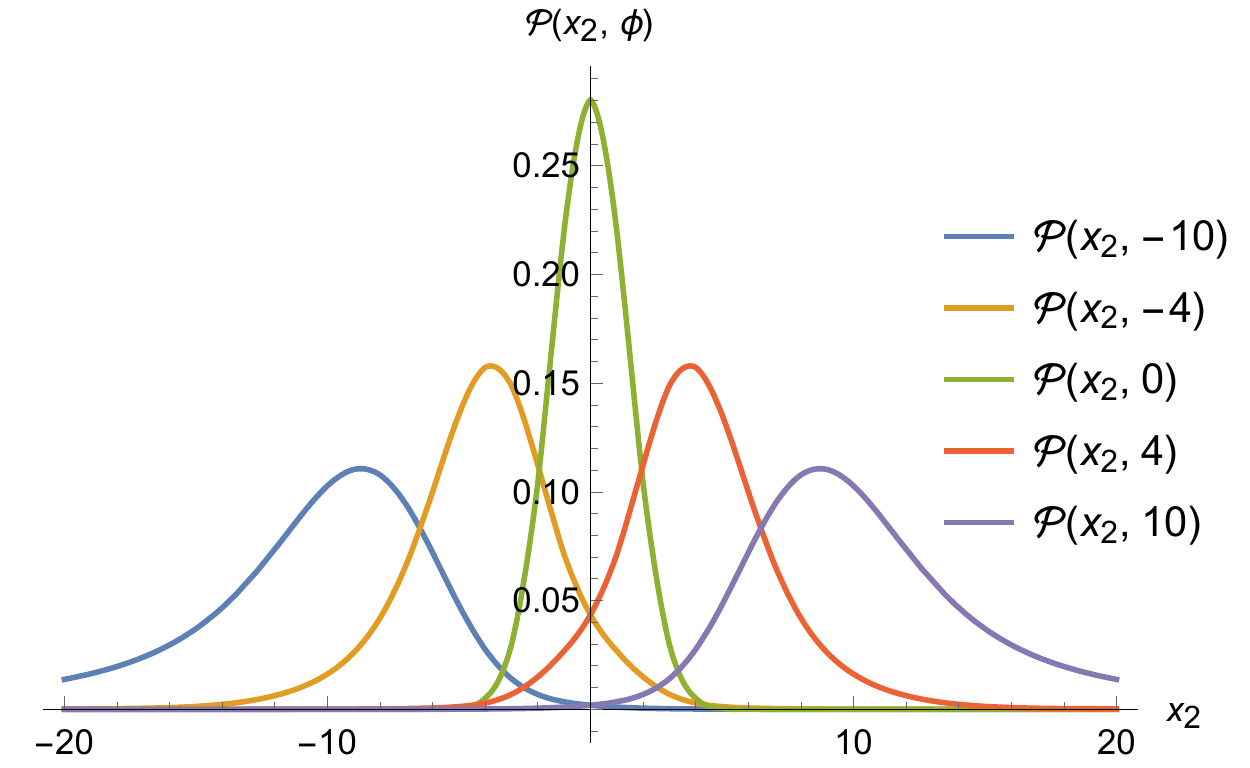}
	\end{minipage}
	\qquad\qquad\qquad
	\begin{minipage}{3.4cm}
		\includegraphics[width=1.5\linewidth]{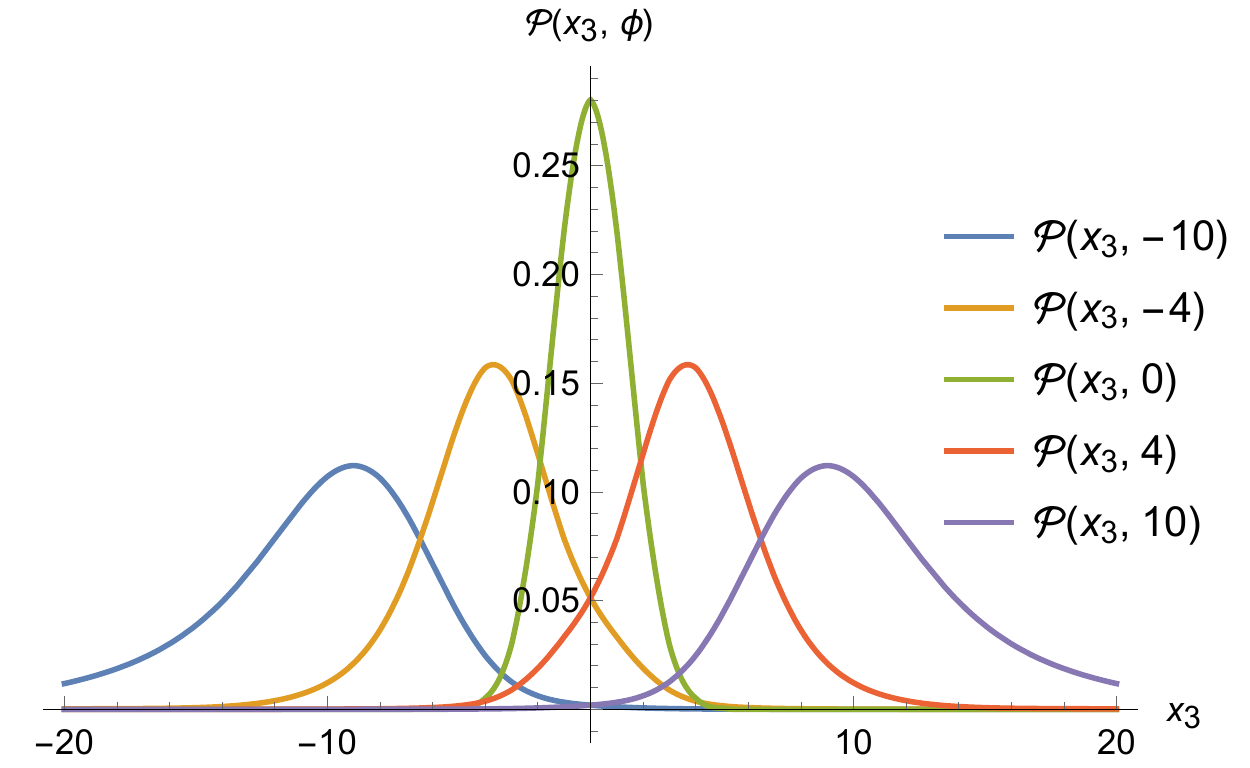}
	\end{minipage}
	\qquad\qquad\qquad\quad
	\caption{The normalized sections $\mathcal{P}(x_i,\phi)$ (with $x_j\sim m_j\phi$ and $x_k\sim m_k\phi$) are shown in sequence for $i=1,2,3$ respectively at different times (here $\bar{k}_1=\bar{k}_2=\bar{k}_3=0, \sigma_{k_1}=\sigma_{k_2}=\sigma_{k_3}=1/2$). Their spreading behavior over time is evident together with the gaussian-like shape.}
	\label{qprob}
\end{figure*}

\begin{figure*}
	\begin{minipage}{3.4cm}
		\includegraphics[width=1.5\linewidth]{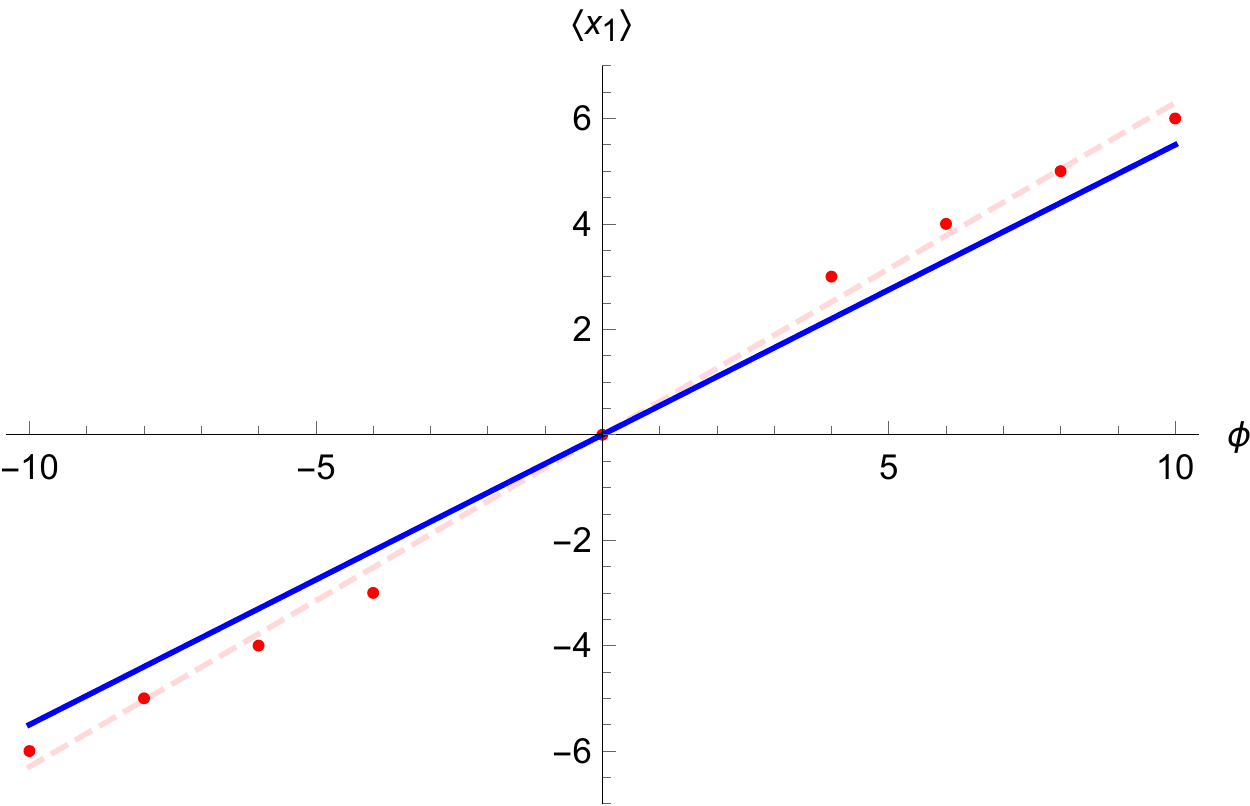}
	\end{minipage}
	\qquad\qquad\qquad
	\begin{minipage}{3.4cm}
		\includegraphics[width=1.5\linewidth]{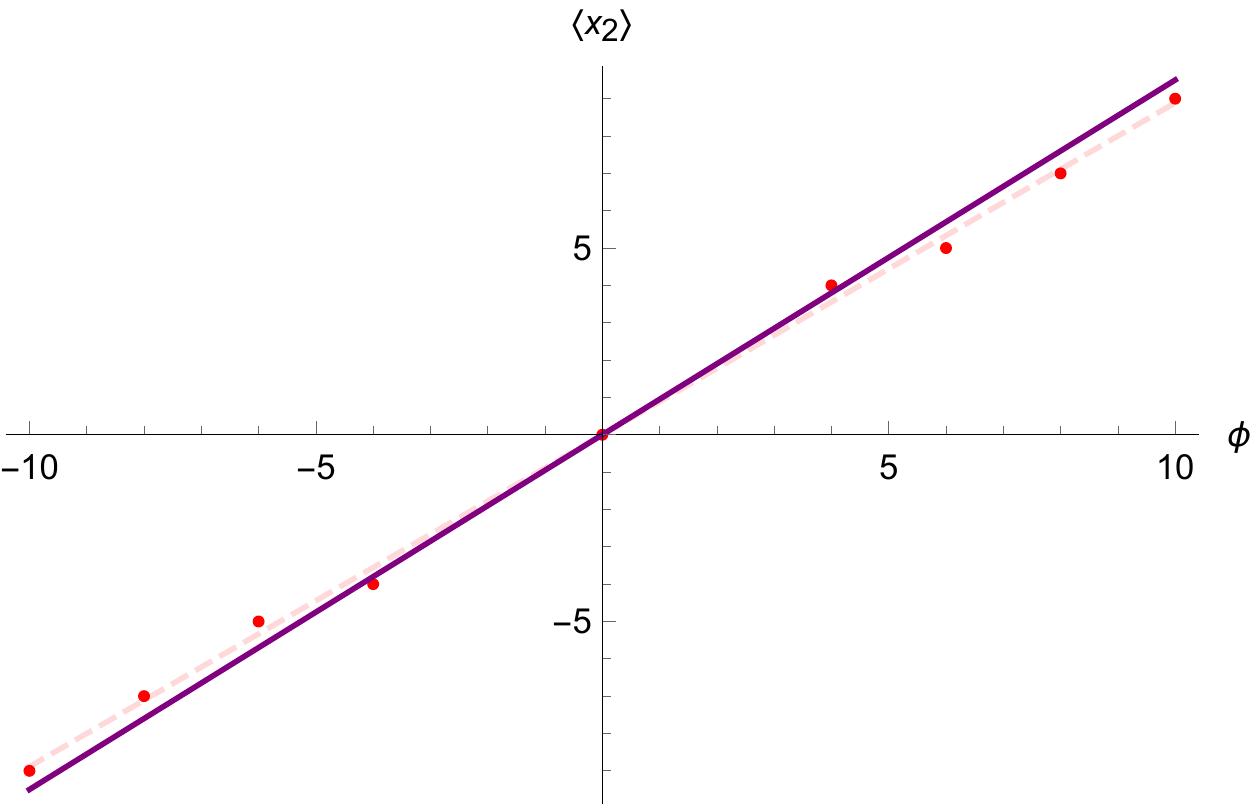}
	\end{minipage}
	\qquad\qquad\qquad
	\begin{minipage}{3.4cm}
		\includegraphics[width=1.5\linewidth]{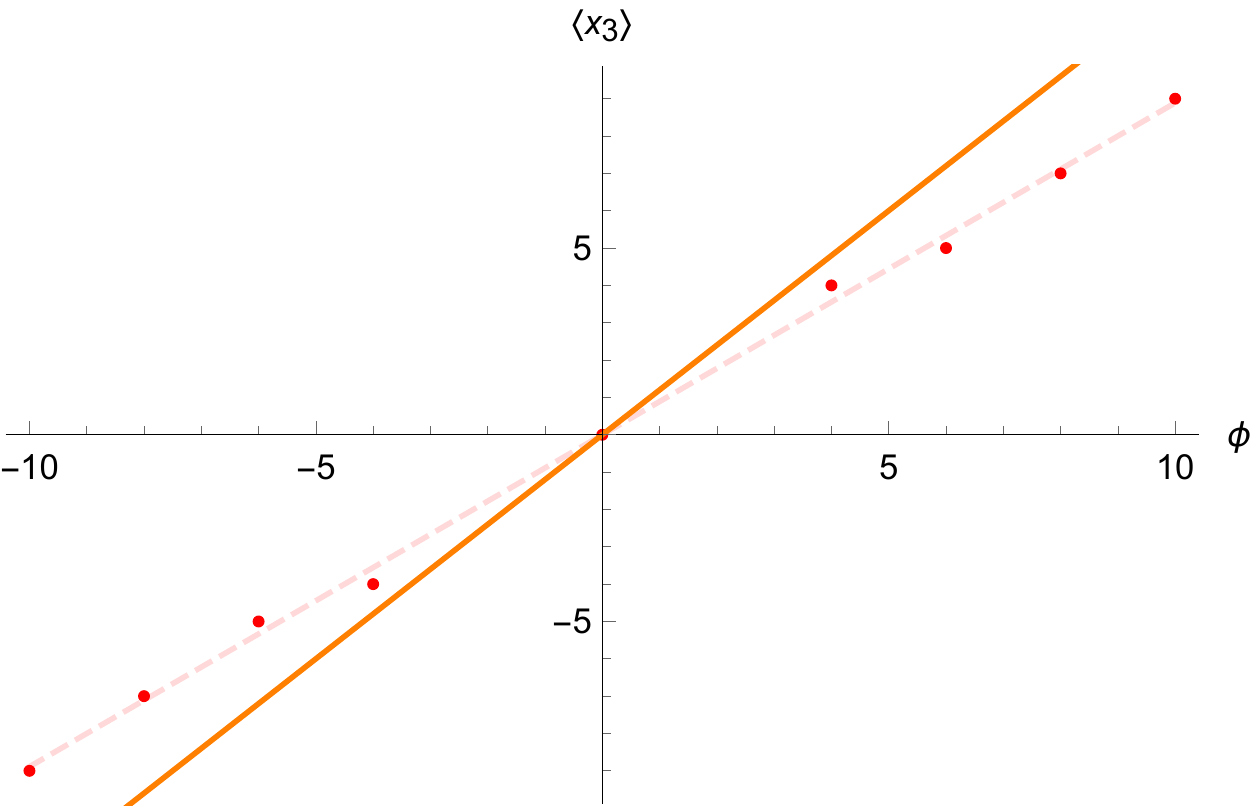}
	\end{minipage}
	\qquad\qquad\qquad\quad
	\caption{The three pictures show the position of the peaks of $\mathcal{P}(x_i,\phi)$ for $i=1,2,3$ respectively in function of time $\phi$ (red dots). The resulting fitting functions (red dashed straight lines) overlap the semiclassical trajectories (continuous lines) with a confidence level of three standard deviations (we have set $m_1=0.55,\,m_2=0.95,\,m_3=1.20$).}
	\label{qpeak}
\end{figure*}
Now, we recall the semiclassical scalar constraint written in the Ashtekar variables \eqref{eqn:Hpoly} with the aim of performing an ADM reduction of the variational principle:
\begin{equation}
	\label{scalar}
	p_{\phi}^{2}-\Theta=0\,,
\end{equation}
where
\begin{align}
	&\Theta={2\over\kappa\gamma^{2}}\Big[ {\sin(\mu_1c_1)p_1\sin(\mu_2c_2)p_2\over\mu_1\mu_2}+\\&+{\sin(\mu_1c_1)p_1\sin(\mu_3c_3)p_3\over\mu_1\mu_3}+{\sin(\mu_2c_2)p_2\sin(\mu_3c_3)p_3\over\mu_2\mu_3}\Big]\,.
\end{align}
After choosing the scalar field $\phi$ as the temporal parameter, we derive the ADM-reduced Hamiltonian by solving the scalar constraint \eqref{scalar} with respect to the momentum associated to the scalar field:
\begin{equation}
	p_{\phi}\equiv\mathcal{H}_{ADM}=\sqrt{\Theta}\,,
	\label{ADM}
\end{equation}
where we choose the positive root in order to guarantee the positive character of the lapse function (see \eqref{shiftf}).
Exploiting this procedure, the Wheeler-DeWitt equation can be rewritten in the form of a Schr\"odinger one by promoting the ADM-reduced Hamiltonian to a quantum operator:
\begin{equation}
	\label{eqn:schro}
	-i\partial_{\phi}\Psi=\sqrt{\hat{\Theta}}\Psi\,,
\end{equation}
where the operator $\sqrt{\hat{\Theta}}$, that we assume well-defined, can be written as
\begin{align}
	\label{sqo}
	\sqrt{\hat{\Theta}}= \Big[-{2\over\kappa\gamma^{2}}\Big( \partial_{x_{1}}\partial_{x_{2}}+\partial_{x_{1}}\partial_{x_{3}}+ \partial_{x_{2}}\partial_{x_{3}}\Big)\Big]^{1/2}\,,
\end{align}
thanks to the substitution \eqref{xiS}. The associated probability density is
\begin{equation}
	\mathcal{P}(\vec x,\phi)=\Psi^{*}(\vec x,\phi)\Psi(\vec x,\phi)\,,
	\label{P}
\end{equation}
where 
\begin{align}
	\Psi(\vec{x},\phi)=A\iiint_{-\infty}^{\infty}dk_1	\,dk_2\,dk_3\,\prod_{i=1}^3\exp\bigg[-{\frac{(k_i-\bar{k}_i)^2}{2\sigma_{k_i}^2}}\bigg]\times\nonumber
	\\ \times  e^{i(k_1x_1+k_2x_2+k_3x_3+\sqrt{\frac{2}{\kappa\gamma^2}|k_1k_2+k_1k_3+k_2k_3|}\phi)}
	\label{packbis} 
\end{align}
and $A$ is fixed by the normalization. Indeed, as demonstrated in \cite{SQO,Puzio} the square root operator \eqref{sqo} possesses plane wave solutions with only positive eigenvalues, so it is always possible to deal with a localized solution and in this sense the problem of non-locality is avoided. Actually, the non-locality is reflected in the fact that the pseudo-differential operator \eqref{sqo} has an infinite order of derivatives and this affects also the absence of a continuity equation for the probability density \eqref{P} that is only globally but non-locally conserved. Hence, we have discarded the covariant formulation of the Wheeler-DeWitt equation in favor of a Schr\"odinger-like one in order to define the probability density \eqref{P} that is positive everywhere by definition and whose spatial integral remains constant through time. This way, we can analyze the quantum dynamics of \eqref{packbis}, in order to verify the consistency between the information carried by the quantum wave packet of the Bianchi I model and the semiclassical solutions provided in Sec. \ref{sem1} when the Ashtekar variables are considered. 

In Fig. \ref{qprob} some different sections of the probability density $\mathcal{P}$ at different values of $\phi$ are shown in order to analyze how its shape and its maximum changes over time. In addition, the present sections have been obtained by fixing two of the three coordinates through the values that they assume in the semiclassical trajectories. More specifically, by combining the analytical solutions for the connections $c_i$ (see \eqref{sysp}) and the relation \eqref{xiS}, we obtain that $x_i$ verifies a pure linear behavior in function of $\phi$. In particular, we have that
\begin{equation}
	\begin{aligned}
&x_i(\phi)\sim\frac{\bar{p}_j/\mu_j+\bar{p}_k/\mu_k}{\gamma p_\phi}\phi=m_i\phi\,
\end{aligned}
\end{equation}
for $i,j,k=1,2,3\,i\neq j\neq k$. We want to remark that this analysis based on the probability density sections is justified by the semiclassical decoupling of the dynamics along the three directions (see Sec.$\,$\ref{sem1}).

As we can see from Fig. \ref{qprob}, the normalized quantum distributions of $x_1,x_2,x_3$ respectively are shown in sequence and their spreading behavior over time is evident, as well as their gaussian-like shape. Also, in Fig. \ref{qpeak} the position of the peaks of $\mathcal{P}$ are represented by the red dots that have also been fitted by means of a linear interpolation. The resulting fitting functions are represented by the red dashed straight lines, whereas the semiclassical trajectories by the continuous straight ones. In particular, the slope of the fitting straight lines is consistent with the semiclassical one $m_i$ ($i=1,2,3$) with a confidence level of the order of three standard deviations for all the three coordinates. Therefore, we can conclude that there is a good correspondence between the quantum behavior of the Universe wave function and the solutions of the semiclassical dynamics. Furthermore, this feature of our analysis in the Ashtekar variables suggests the presence of a bouncing dynamics with non-universal properties also at a quantum level when the polymer paradigm is fully implemented. 

Now, let us consider the proper volume set and compute the quantum mean value of the volume operator and its standard deviation. The aim of this analysis is to investigate the consequences of the spreading behavior of the Bianchi I wave packet on the quantum fluctuations. Recalling the Hamiltonian constraint \eqref{HV}, we write the Schr\"{o}dinger-like equation in the volume variables
\begin{equation}
	\label{eqn:schro2}
-i\partial_{\phi}\Psi=\sqrt{\hat{\Xi}}\Psi\,,
\end{equation}
where $\hat{\lambda}_{1,2}=-i\partial_{\eta_{1,2}}$, $\hat{v}=-i\partial_{\eta_3}$ and the ADM-reduced Hamiltonian can be written as
\begin{equation}
\sqrt{\hat{\Xi}}=\Big[-{1\over2\kappa\gamma^{2}}\Big( \partial_{x_{1}}\partial_{x_{2}}+2\partial_{x_{1}}\partial_{x_{3}}+2 \partial_{x_{2}}\partial_{x_{3}}+3\partial^2_{x_{3}}\Big)\Big]^{1/2}\,
\end{equation}
thanks to the substitution $x_i=\ln\big[{\tan\big({\frac{\mu_i \eta_i}{2}}}\big)\big]+\bar{x}_i$. The mean value of the volume operator can be computed as
\begin{equation}
\int_{-\infty}^{+\infty}d\vec{x}\,\Psi^*(\vec{x},\phi)\hat{v}\Psi(\vec{x},\phi)\,,
\label{vmedio}
\end{equation}
where
\begin{equation}
	\label{packbis2}
\begin{aligned}
	\Psi(\vec{x},\phi)=A\int_{-\infty}^{\infty}dk_3\,&e^{-\frac{(k_3-\bar{k}_3)^2}{2\sigma^2}}e^{i(\bar{k}_1x_1+\bar{k}_2x_2+k_3x_3)}\times \\ \times &e^{i\sqrt{\frac{1}{2\kappa\gamma^2}\big|\bar{k}_1\bar{k}_2+2\bar{k}_1k_3+2\bar{k}_2k_3+3k_3^2\big|}\phi}
\end{aligned}
\end{equation}
is the Universe wave packet, that is constructed through the superposition of the plane wave solutions of \eqref{eqn:schro2} by means of a gaussian coefficient on $k_3$, and $A$ is fixed by the normalization. Moreover, in order to quantify the consistency between the quantum behavior and the semiclassical one, we have to consider the standard deviation, i.e.
\begin{equation}
	\Delta\hat{v}=\sqrt{\braket{\hat{v}^2}-\braket{\hat{v}}^2}\,.
\end{equation}
In order to simplify the numerical calculation of the volume mean value, fixed values have been chosen for $k_1,k_2$. In other words, we have considered Gaussian distributions for $k_1,k_2$ so narrow that they can be reasonably approximated with a Dirac delta. In \eqref{vmedio} we remark that we have used the substitution from $\eta_3$ to $x_3$ to write the $\hat{v}$ operator in the $x$-representation. 
\begin{figure}
	\centering
	\includegraphics[width=0.7\linewidth]{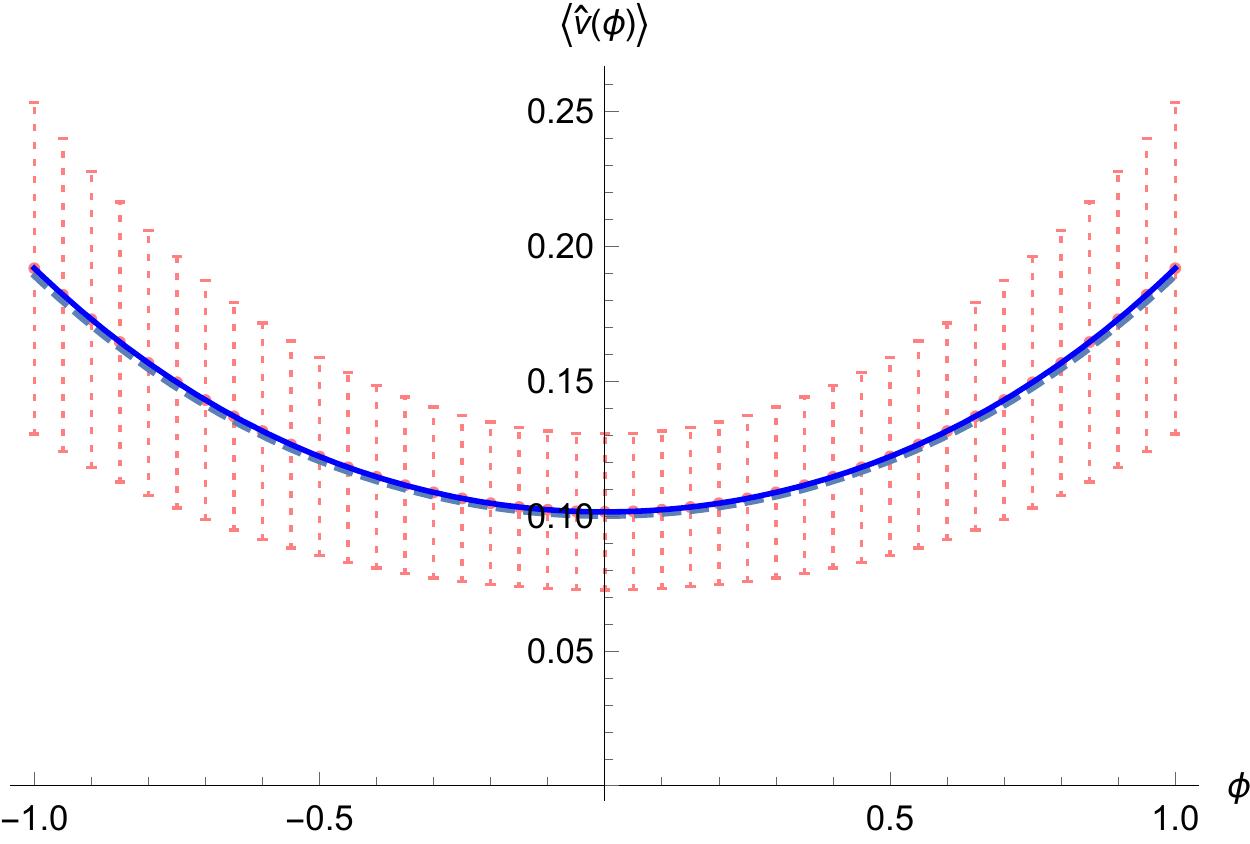}
	\caption{The mean value of the volume operator  $\braket{\hat{v}}$ as function of $\phi$ (pink dots), the interpolation function (continuous blue line) and the semiclassical trajectory (dashed gray line). The error bars (pink dashed vertical lines) are centered in the mean values and have a length of $2\Delta\hat{v}$. We have set $\kappa=\gamma=1,\sigma=4,\bar{k}_1=11,\bar{k}_2=9,\bar{k}_3=10$.}
	\label{VbarNEW}
\end{figure}
\begin{figure}
	\centering
	\includegraphics[width=0.7\linewidth]{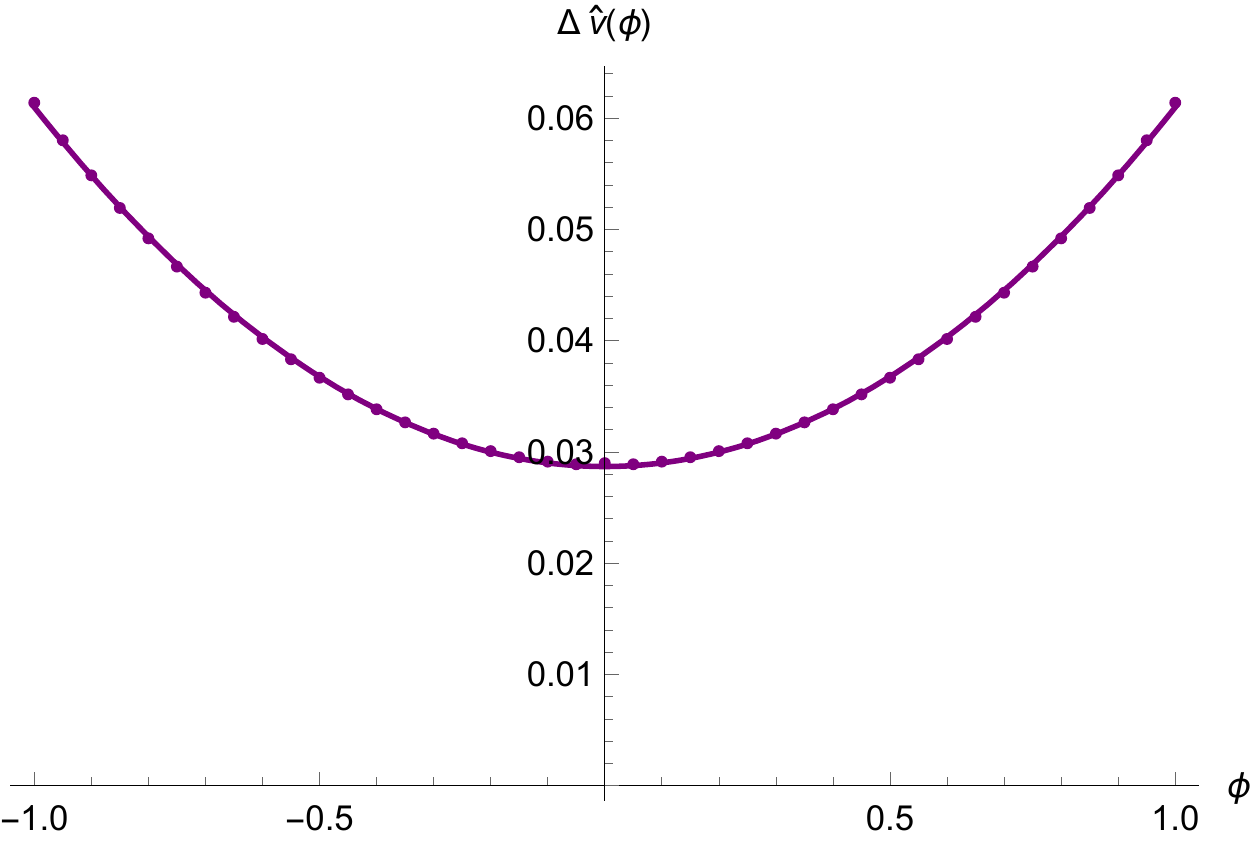}
	\caption{The standard deviation of the volume operator  $\Delta\hat{v}$ as function of $\phi$ (purple dots) and the parabolic interpolation function (continuous purple line). We have set $\kappa=\gamma=1,\sigma=4,\bar{k}_1=11,\bar{k}_2=9,\bar{k}_3=10$.}
	\label{VquadromedioNEW}
\end{figure}

\begin{figure}
	\centering
	\includegraphics[width=0.7\linewidth]{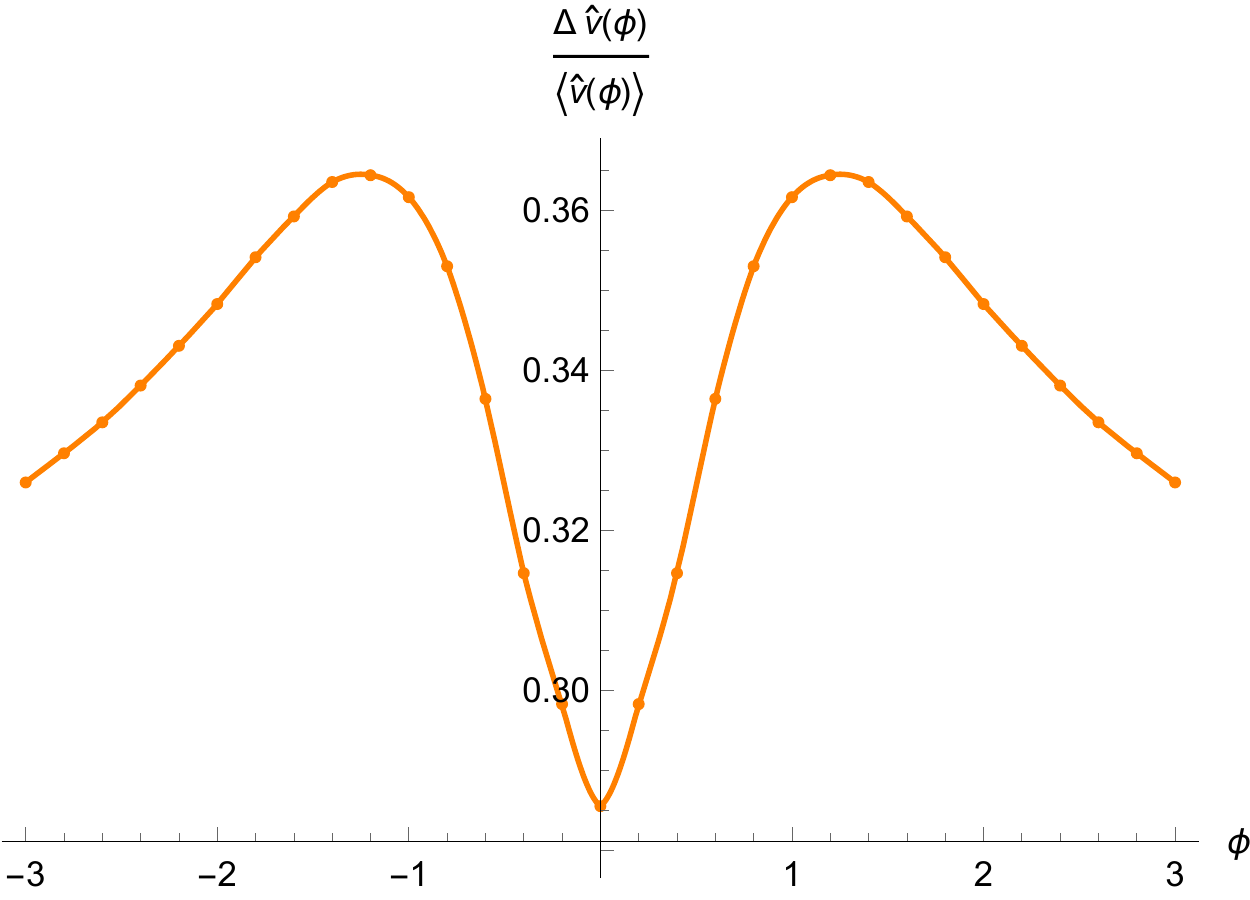}
	\caption{The relative error $\Delta\hat{v}/\braket{\hat{v}}$ as function of $\phi$ (orange dots) and the interpolation function (continuous orange line).}
	\label{Vrel}
\end{figure}

In Fig. \ref{VbarNEW} the dots represent the mean value of $\braket{\hat{v}}$ at fixed times and the continuous line is the resulting interpolation function, that well follows the semiclassical trajectory for the Universe volume \eqref{vphii} (dashed line). On the other hand, the spreading phenomenon is evident in the error bars (see Fig. \ref{VbarNEW}). Indeed, as we can see from Fig. \ref{VquadromedioNEW}, $\Delta\hat{v}$ grows with $\phi$ in a non-linear way. In order to derive the interpolation function, the dots representing the value of $\Delta\hat{v}$ at fixed times have been fitted, obtaining a quadratic relation with time, i.e. $\Delta\hat{v}\sim a+b\phi^2$. This means that the Bianchi I wave packet remains localized only in the neighborhood of $\phi=0$, where the consistency with the semiclassical solution is meaningful. So, referring to Fig. \ref{VbarNEW} we can infer that near the critical point there is a good correspondence with the semiclassical solution $v(\phi)$, but as $\phi$ grows the standard deviation becomes so relevant that the comparison with the semiclassical behavior is meaningless. We notice that the highly-localization at the Big Bounce is a result of how the Bianchi I wave packet has been constructed and does not have a physical meaning. Indeed, the Universe wave function can be initially localized at any point of the configurational space, actually also in correspondence of the Bounce. This arbitrariness corresponds to the possibility of admitting the quasi-classical limit near the Bounce, as shown in Fig. \ref{VbarNEW}. However, the unlimited growth of $\Delta\hat{v}$, i.e. the spreading of the wave packet, questions the possibility to extrapolate  the quasi-classical limit at any time. Actually, in Fig. \ref{Vrel} the behavior of the relative error $\Delta\hat{v}/\braket{\hat{v}}$ shows that the Universe wave packet has a maximum  localization at the Bounce, then the quantum fluctuations grows until it is reached a region of minimum localization, after which the relative error decreases. In conclusion, this analysis highlights that, even if the validity of the quasi-classical limit is not prevented in this picture, the semiclassical characterization of the Big Bounce for an anisotropic model (in which the spreading phenomenon is present) does not give a complete picture of the Universe dynamics since the role of the quantum fluctuations can not be neglected.

\section{CONCLUDING REMARKS\label{con}}

In this paper we analyzed the semiclassical and quantum dynamics of
the Bianchi I cosmology in the presence of a massless scalar field as viewed in the polymer paradigm, which is able to introduce a discrete nature in the behavior of the configurational variables.
Our study is focused on the comparison between the dynamical features emerging
when the natural Ashtekar connections or the volume-like variables are
implemented.
More specifically, the anisotropic character of the Bianchi I model has allowed us to adopt two different sets of volume-like variables,
one \cite{ashtekar2009} (see also \cite{M, Ant}) that corresponds to adopting
the Universe volume itself (de facto the product of all the three scale factors)
and the other \cite{szulc} which consists of a set of three anisotropic volume-like
variables. It is important to stress that in the case of an isotropic
Universe it would reduce to a unique choice, in particular that one at the ground of the so-called improved LQC Hamiltonian
\cite{AshtekarI}.

On the semiclassical level, i.e. when we deal with a modified dynamics
due to a redefined conjugate momentum, we provided a detailed study in all the three
sets of variables. In LQC this analysis is justified by the considerations developed in \cite{Rovelli2,B1,B2}, so the present polymer analysis is expected to confirm the possibility to have the quasi-classical limit near the Bounce. In the theoretical framework traced above we clarified how the emergence
of a universal Bounce in the past Universe evolution (i.e. fixed by fundamental constants and parameters only) is guaranteed only using the Universe volume itself as a configurational coordinate defined on the polymer lattice. This analysis is also supported by the derivation of the exact polymer-modified Friedmann equation for the Bianchi I model in a convenient form. This way, the proper expression of the critical energy density for the anisotropic Universe has been provided, giving a complete picture of the Big Bounce features. In this respect, it is interesting to highlight that choosing anisotropic volume-like variables (i.e. coordinates with the geometrical dimensions of a volume) does not ensure the presence of an intrinsic cut-off, even if they reduce to the Universe volume in the isotropic limit.
This issue could have some impact in the attempt to clarify which
property of the configurational space
is associated to a bouncing cosmology when PQM
is applied (for a discussion of this
question see \cite{EBianchiIX, EFG}). Similarly, when we adopted the Ashtekar connections (adapted to the Bianchi
I model) we demonstrated that the
Big Bounce is not a universal feature of the cosmological model, since it depends on the initial conditions on the dynamics.
Therefore, if we regard as preferable the use of the Ashtekar connections in accordance with the paradigm of LQG, we can conclude that the polymer quantum dynamics of a Bianchi I model is always associated to a Bounce, but the  critical energy density does not take a fixed cut-off value. Moreover, we have also shown that it is possible to recover a dynamical equivalence between the Ashtekar set and the anisotropic volume-like one. In particular, by considering the polymer parameter as depending on the configurational coordinates when performing the canonical change of variables, in the two formulations we obtain the same dynamics of that set in which the polymer parameter is constant. Clearly, the dynamical equivalence should be demonstrated also in the full quantum picture, where it involves the non-trivial implementation of a translational operator depending on the coordinates. However, the intrinsic one-dimensional character of the polymer representation prevents the generalization of the proposal made in the FLRW case \cite{EFG} to a six-dimensional phase space (as that of the Bianchi I model) when the momenta directions are mixed. As a consequence, it is not possible to demonstrate the equivalence between the proper volume formulation and the Ashtekar or the anisotropic volume-like one, not even at a semiclassical level. In this sense, developing a tree-dimensional picture of the polymer formulation would be useful to understand what assumptions make it possible to reproduce a universal bouncing dynamics also in the Ashtekar variables. This would also highlight the strong parallelism between the $\mu_0$ and $\bar{\mu}$ schemes of LQC to the polymerization of the Ashtekar and proper volume variables for the Bianchi I model, as demonstrated in the isotropic case in \cite{EFG}.

Then, we performed the canonical quantization of the polymer-modified model in the Ashtekar connections paradigm and in the proper volume variables to study the behavior of the Universe wave packet.
Actually, by performing an ADM reduction of the
variational problem we passed to a Schr\"{o}dinger-like
representation, in order to avoid the issue regarding the sign of the Klein-Gordon probability density that would be emerged in the Wheeler-DeWitt approach
(in this respect, see \cite{shell,rosenstein1985probability}). Regarding the Ashtekar representation, by following the dynamics of the probability density peak we saw that it has common features with the corresponding semiclassical trajectories and this
leads us to claim that also on a quantum level the use of Ashtekar variables provides a bouncing cosmology whose critical energy density depends on the distribution of the quantum numbers characterizing the Universe wave packet. Furthermore, in the proper volume representation we have computed the quantum mean value of the volume operator, showing the consistency with the semiclassical Universe volume trajectory but also the non-linear behavior of the standard deviation. This result is due to the spreading behavior of the Bianchi I wave packet over time, as shown in the quantum analysis in the Ashtekar variables, differently from what happens in the isotropic case. So, even if the relative error is bounded and the quasi-classical limit is not prevented, the relevance of the quantum fluctuations makes the semiclassical analysis near the Bounce a qualitative but not completely satisfactory approach in the case of anisotropic cosmological models \cite{QBB}.

By concluding, thanks to the relevant role that the Bianchi I model plays in constructing the behavior towards the singularity of a Bianchi IX cosmological model (and
hence of a generic inhomogeneous Universe \cite{BKL82,PC}), we claim that also in more general cosmological models our results suggest the presence of a Big Bounce with the same features traced here, especially when the Ashtekar connections are concerned
\cite{ABKL}.
	
\nocite{*}
\bibliography{apssampSilvia.bib}
		
\end{document}